\documentclass[twocolumn]{article}

\usepackage[utf8]{inputenc}
\usepackage{amsmath,amssymb}
\usepackage{bm}
\usepackage{bbold}
\usepackage{mathrsfs}
\usepackage{booktabs}
\usepackage{graphicx}
\usepackage{adjustbox}

\usepackage{widetext}

\usepackage[left=3cm,right=3cm,top=3cm,bottom=3cm]{geometry}
\usepackage{enumerate}

\usepackage{aas_macros}
\usepackage{authblk}
\usepackage{hyperref}
\usepackage{color}

\newcommand\T{\rule{0pt}{2.9ex}}       
\newcommand\B{\rule[-1.4ex]{0pt}{0pt}} 

\title{Some Recent Results on Neutrino Oscillations in Hypercritical Accretion}
\author[1,5]{J. D. Uribe\thanks{juan.uribe@icranet.org}}
\author[1,2,3,4]{J. A. Rueda\thanks{jorge.rueda@icra.it}}

\affil[1]{ICRANet, P.zza della Repubblica 10, I--65122 Pescara, Italy}
\affil[2]{ICRANet-Ferrara, Dipartimento di Fisica e Scienze della Terra, Universit\`a degli Studi di Ferrara, Via Saragat 1, I--44122 Ferrara, Italy}
\affil[3]{Dipartimento di Fisica e Scienze della Terra, Universit\`a degli Studi di Ferrara, Via Saragat 1, I--44122 Ferrara, Italy}
\affil[4]{INAF, Istituto di Astrofisica e Planetologia Spaziali, Via Fosso del Cavaliere 100, 00133 Rome, Italy}
\affil[5]{Dipartimento di Fisica, Sapienza Universit\`a di Roma, P.le Aldo Moro 5, I--00185 Rome, Italy}

\date{\today}

\begin{document}

\maketitle

\begin{abstract}
	The study of neutrino flavour oscillations in astrophysical sources has been boosted in the last two decades thanks to achievements in experimental neutrino physics and observational astronomy. We here discuss two cases of interest in the modelling of short and long gamma-ray bursts (GRBs): hypercritical, i.e. highly super-Eddington spherical/disk accretion onto a neutron star (NS)/black hole (BH). We show that in both systems the ambient conditions of density and temperature imply the occurrence of neutrino flavour oscillations, with a relevant role of neutrino self-interactions.	
\end{abstract}

\section{Introduction}\label{sec1}

The occurrence of neutrino flavour oscillations has been, undoubtedly, experimentally demonstrated~\cite{deSalas:2017kay}. Of special interest here is that, it has become clear in recent years that for the analysis of neutrino oscillations in matter, e.g. the Mikheyev-Smirnov-Wolfenstein (MSW) effect~\cite{1978PhRvD..17.2369W,Mikheyev1986}, refractive effects of neutrinos on themselves, due to the neutrino self-interaction potential, are essential.

Over the last two decades, the achievements of experimental neutrino physics and the constant development of observational astronomy, have caused an increasing interest in the study of the occurrence of neutrino flavour oscillations in astrophysical sources. Although the bulk of astrophysical analyses has been limited to supernova (SN) neutrinos, flavour oscillations may also occur in other relativistic astrophysics sources. In particular, as we are showing here, this phenomenon is expected to occur in known scenarios of short- and long-duration GRBs.

The emergent picture of GRBs is that both, short-duration and long-duration GRBs, originate in binary systems (see, e.g., \cite{2016ApJ...832..136R}). Short bursts are associated with mergers of NS-NS and/or NS-BH binaries. For this case, the role of neutrino-antineutrino ($\nu\bar\nu$) annihilation leading to the formation of an electron-positron plasma ($e^{-}e^{+}$) has been introduced \cite{1992ApJ...395L..83N} (for general relativistic effects, see \cite{2002ApJ...578..310S}). For long bursts, it has been introduced binary progenitor composed of a carbon-oxygen star (CO$_{\rm core}$) and a companion NS~\cite{2012ApJ...758L...7R,2014ApJ...793L..36F}. These binaries can form in an evolutionary path including a first SN explosion, common-envelope phases, tidal interactions and mass loss \cite{2015PhRvL.115w1102F}. The GRB is expected to occur when the binary experiences the second SN, i.e. the one of the CO$_{\rm core}$. Part of the ejected matter produces a hypercritical accretion (i.e. highly super-Eddington) process onto the NS companion. The NS then reaches its critical mass for gravitational collapse, hence forming a rotating BH \cite{2015ApJ...812..100B,2016ApJ...833..107B}. These systems have been called binary-driven hypernovae (BdHNe), and they lead to a variety of observable emissions from the X-rays all the way to high-energy gamma-rays (see e.g. \cite{2019ApJ...886...82R,2019Univ....5..110R} for details).

As we are showing below, a key ingredient in the above systems is a copious emission of neutrinos during the hypercritical accretion. The high neutrino and matter density involved suggests that a study of neutrinos oscillations may lead to new neutrino physics in these sources. Our aim here is to compile the main results of neutrino oscillations in the physical conditions expected in the above scenarios of GRBs.

\section{Neutrino Oscillations}\label{sec2}

To study the flavour evolution of neutrinos within a particular system, a Hamiltonian governing neutrino oscillation must be set up. The relative strength of the potentials appearing in such Hamiltonian depends on four elements: geometry, mass content, neutrino content and neutrino mass hierarchy. Geometry refers to the nature of net neutrino fluxes and possible gravitational effects. Mass and neutrino content refers to the distribution of leptons of each flavour $(e,\mu,\tau)$ present in the medium. Finally, mass hierarchy refers to the relative values of the masses $m_{1},m_{2},m_{3}$ for each neutrino mass eigenstates. The equations that govern the evolution of an ensemble of mixed neutrinos are the quantum Liouville equations
\begin{subequations}
\begin{gather}
i\dot{\rho}_{\mathbf{p}} = [H_{\mathbf{p}},\rho_{\mathbf{p}}]\\
i\dot{\bar{\rho}}_{\mathbf{p}} = [\bar{H}_{\mathbf{p}},\bar{\rho}_{\mathbf{p}}]\end{gather}\label{eq:Liouville}\end{subequations}
The Hamiltonian is (see, e.g.,~\cite{2018ApJ...852..120B,2019arXiv190901841U} and references therein)

\begin{widetext}
\begin{subequations}
\begin{align}
\mathsf{H}_{\mathbf{p},t}&=\Omega_{\mathbf{p},t}+\sqrt{2}G_{F}\!\!\int\!\!\left( l_{\mathbf{q},t}-\bar{l}_{\mathbf{q},t}\right)\left( 1-\mathbf{v}_{\mathbf{q},t}\cdot\mathbf{v}_{\mathbf{p},t} \right)\frac{d^3\mathbf{q}}{\left(2\pi\right)^3} \nonumber \\
&\qquad\qquad\qquad\qquad\qquad + \sqrt{2}G_{F}\!\!\int\!\!\left( \rho_{\mathbf{q},t}-\bar{\rho}_{\mathbf{q},t}\right)\left( 1-\mathbf{v}_{\mathbf{q},t}\cdot\mathbf{v}_{\mathbf{p},t} \right)\frac{d^3\mathbf{q}}{\left(2\pi\right)^{3}}\\
\mathsf{\bar{H}}_{\mathbf{p},t}&=-\Omega_{\mathbf{p},t}+\sqrt{2}G_{F}\!\!\int\!\!\left( l_{\mathbf{q},t}-\bar{l}_{\mathbf{q},t}\right)\left( 1-\mathbf{v}_{\mathbf{q},t}\cdot\mathbf{v}_{\mathbf{p},t} \right)\frac{d^3\mathbf{q}}{\left(2\pi\right)^3} \nonumber \\
&\qquad\qquad\qquad\qquad\qquad + \sqrt{2}G_{F}\!\!\int\!\!\left( \rho_{\mathbf{q},t}-\bar{\rho}_{\mathbf{q},t}\right)\left( 1-\mathbf{v}_{\mathbf{q},t}\cdot\mathbf{v}_{\mathbf{p},t} \right)\frac{d^3\mathbf{q}}{\left(2\pi\right)^{3}}
\end{align}\label{eq:FullHam}\end{subequations}\end{widetext}
In these equations $\rho_{\mathbf{p}}$ ($\bar{\rho}_{\mathbf{p}}$) is the matrix of occupation numbers $(\rho_{\mathbf{p}})_{ij}=\langle a^{\dagger}_{j}a_{i}\rangle_\mathbf{p}$ for neutrinos ($(\bar{\rho}_{\mathbf{p}})_{ij}=\langle \bar{a}^{\dagger}_{i}\bar{a}_{j}\rangle_\mathbf{p}$ for antineutrinos), for each momentum $\mathbf{p}$ and flavours $i,j$. The diagonal elements are the distribution functions $f_{\nu_{i}\left(\bar{\nu}_{i}\right)}\left(\mathbf{p}\right)$ such that their integration over the momentum space gives the neutrino number density $n_{\nu_{i}}$ of a determined flavour $i$. The off-diagonal elements provide information about the \emph{overlapping} between the two neutrino flavours. $\Omega_{\mathbf{p}}$ is the matrix of vacuum oscillation frequencies, $l_{\mathbf{p}}$ and $\bar{l}_{\mathbf{p}}$ are matrices of occupation numbers for charged leptons built in a similar way to the neutrino matrices, and $\mathbf{v}_{\mathbf{p}}=\mathbf{p}/ p$ is the velocity of a particle with momentum $\mathbf{p}$ (either neutrino or charged lepton). Since the matter in the accretion zone is composed by protons, neutrons, electrons and positrons, $\nu_e$ and $\bar\nu_e$ interact with matter by both charged and neutral currents, while $\nu_\mu$, $\nu_\tau$, $\bar\nu_\mu$ and $\bar\nu_\tau$ interact only by neutral currents. Therefore, the behavior of these states can be clearly divided into electronic and non-electronic allowing us to use the two-flavour approximation. Within this approximation, $\rho$ in Eq.~(\ref{eq:Liouville}) can be written in terms of Pauli matrices and the polarization vector $\mathsf{P}_\mathbf{p}$ as:
\begin{equation}
\small
\rho_{\mathbf{p}}=\left(
 \begin{array}{cc}
  \rho_{ee} & \rho_{ex}\\
  \rho_{xe} & \rho_{xx}\\
   \end{array}\right)_{\mathbf{p}}
   =
 \frac{1}{2}\left(f_{\mathbf{p}}\mathbb{1} +\mathsf{P}_\mathbf{p} \cdot \vec \sigma\right),
	\label{eq:expansion of rho}
\end{equation}
where $f_{\mathbf{p}}={\rm Tr}[\rho_{\mathbf{p}}]=f_{\nu_e}(\mathbf{p})+f_{\nu_x}(\mathbf{p})$ is the sum of the distribution functions for $\nu_e$ and $\nu_x$. Note that the $z$ component of the polarization vector obeys
\begin{equation}
\mathsf{P}^{z}_{\mathbf{p}} =f_{\nu_e}(\mathbf{p})-f_{\nu_x}(\mathbf{p}).
\label{eq:pzeta1}
\end{equation}

Hence, this component tracks the fractional flavour composition of the system and appropriately normalizing $\rho_{\mathbf{p}}$ allows to define a survival and mixing probability

\begin{subequations}
\begin{gather}
P_{\nu_{e} \leftrightarrow \nu_{e}} = \frac{1}{2}\left( 1 + \mathsf{P}^{z}_{\mathbf{p}} \right),\\
P_{\nu_{e} \leftrightarrow \nu_{x}} = \frac{1}{2}\left( 1 - \mathsf{P}^{z}_{\mathbf{p}} \right).
\end{gather}\label{eq:survprobability1}
\end{subequations}

On the other hand, the Hamiltonian can be written as a sum of three interaction terms:
\begin{equation}
\mathsf{H} = \mathsf{H}_{\mbox{\footnotesize{vac}}} + \mathsf{H}_{\mbox{\footnotesize{m}}} + \mathsf{H}_{\nu\nu}.
\label{neutrinohamiltonian}
\end{equation}
where $\mathsf{H}$ is the two-flavour Hamiltonian. The first term is the Hamiltonian in vacuum~\cite{Qian:1994wh}:
\begin{equation}
\mathsf{H}_{\mbox{\footnotesize{vac}}} =\frac{\omega_\mathbf{p}}{2}
\left(
 \begin{array}{cc}
  -\cos 2\theta & \sin 2\theta\\
  \sin 2\theta & \cos 2\theta \\
   \end{array}\right)
   =\frac{\omega_\mathbf{p}}{2} \mathbf{B}\cdot \vec{\sigma}
	\label{Hvacuum}
\end{equation}
where $\omega_\mathbf{p} = \Delta m^2/2p$, $\mathbf{B}=(\sin2\theta,0,-\cos 2 \theta)$ and $\theta$ is the smallest neutrino mixing angle in vacuum. 

The other two terms in Eqs.~(\ref{eq:FullHam}) are special since they make the evolution equations non-linear. Even though they are very similar, we are considering that the electrons during the accretion form an isotropic gas; hence, the vector $\mathbf{v}_{\mathbf{q}}$ in the first integral is distributed uniformly on the unit sphere and the factor $\mathbf{v}_\mathbf{q}\cdot\mathbf{v}_\mathbf{p}$ averages to zero. After integrating the matter Hamiltonian is given by:
\begin{equation}
\mathsf{H}_{\mbox{\footnotesize{m}}} =
\frac{\lambda}{2}\left(
 \begin{array}{cc}
  1 & 0\\
  0 & -1 \\
   \end{array}\right)
   =\frac{\lambda}{2} \mathbf{L} \cdot \vec{\sigma}
	\label{Hmatter}
\end{equation}
where $\lambda = \sqrt{2}G_{F}\left(n_{e^-} - n_{e^+}\right)$ is the charged current matter potential and $\mathbf{L}=(0,0,1)$.

Such simplification cannot be made with the final term. Since neutrinos are responsible for the energy loss of the infalling material during accretion, they must be escaping the accretion zone and the net neutrino and anti-neutrino flux is non-zero.In this case the factor $\mathbf{v}_\mathbf{q}\cdot\mathbf{v}_\mathbf{p}$ cannot be averaged to zero. At any rate, we can still use Eq.~(\ref{eq:expansion of rho}) and obtain \cite{1992PhLB..287..128P,2016arXiv160704671Z,2016PhRvD..93d5021M}:
\begin{equation}
\mathsf{H}_{\nu\nu} = \sqrt{2}G_{F}\left[ \int\!\! \left(1- \mathbf{v}_{\mathbf{q}}\cdot\mathbf{v}_{\mathbf{p}}\right) \left(\mathsf{P}_\mathbf{q}-\bar{\mathsf{P}}_\mathbf{q}\right)\frac{d^3\mathbf{q}}{\left(2\pi\right)^3}\right]\cdot \vec{\sigma}
\label{Hnunu}
\end{equation}

Introducing every Hamiltonian term in Eqs.~(\ref{eq:Liouville}), and using the commutation relations of the Pauli matrices, we find the equations of oscillation for neutrinos and anti-neutrinos for each momentum mode $\mathbf{p}$:
\begin{widetext}
\begin{subequations}
\begin{gather}
\dot{\mathsf{P}}_\mathbf{p} = \left[ \omega_\mathbf{p} \mathbf{B} + \!\lambda \mathbf{L} + \!\! \sqrt{2}G_{F}\!\!\!  \int\!\! \left(1- \mathbf{v}_{\mathbf{q}}\!\cdot\mathbf{v}_{\mathbf{p}}\right) \left(\mathsf{P}_\mathbf{q}-\bar{\mathsf{P}}_\mathbf{q}\right)\frac{d^3\mathbf{q}}{\left(2\pi\right)^3} \right] \times \mathsf{P}_\mathbf{p}\\
\dot{\bar{\mathsf{P}}}_\mathbf{p} = \left[ -\omega_\mathbf{p} \mathbf{B} + \!\lambda \mathbf{L} + \!\! \sqrt{2}G_{F}\!\!\!  \int\!\! \left(1- \mathbf{v}_{\mathbf{q}}\!\cdot\mathbf{v}_{\mathbf{p}}\right) \left(\mathsf{P}_\mathbf{q}-\bar{\mathsf{P}}_\mathbf{q}\right)\frac{d^3\mathbf{q}}{\left(2\pi\right)^3}\right]\times \bar{\mathsf{P}}_\mathbf{p}.\end{gather}\label{eq:Hnu1}\end{subequations}
\end{widetext}

This set of equations is the starting point of any analysis of neutrino oscillation in an astrophysical system. In the next sections 

\subsection{Neutrino Oscillation in Spherical Accretion}\label{sec2.1}

In the BdHN scenario of GRBs, the SN material first reaches the gravitational capture region of the NS companion, namely the Bondi-Hoyle region. The infalling material shocks as it piles up onto the NS surface forming an accretion zone where it compresses and eventually becomes sufficiently hot to trigger a highly efficient neutrino emission process. Neutrinos take away most of the infalling matter's gravitational energy gain, letting it reduce its entropy and be incorporated into the NS. It was shown in~\cite{2016ApJ...833..107B} that the matter in the accretion zone near the NS surface develops conditions of temperature and density such that it is in a non-degenerate, relativistic, hot plasma state. The most efficient neutrino emission channel under those conditions becomes the electron positron pair annihilation process. The neutrino emissivity can be approximated with a very good accuracy by \cite{2001PhR...354....1Y}.
\begin{widetext}
\begin{equation}
\varepsilon^{m}_{i} \approx \frac{2G^{2}_{F}\left(T\right)^{8+m}}{9\pi^{5}}C^{2}_{+,i}\left[\mathcal{F}_{m+1,0}\left(\eta_{e^{+}}\right)\mathcal{F}_{1,0}\left(\eta_{e^{-}}\right) + \mathcal{F}_{m+1,0}\left(\eta_{e^{-}}\right)\mathcal{F}_{1,0}\left(\eta_{e^{+}}\right)\right]\label{eq:approximationyakovlev}
\end{equation}
\end{widetext}

where $\mathcal{F}_{k,\ell}\left(y, \eta \right)$ are the generalized Fermi functions (see \cite{2018ApJ...852..120B} for details) and $\mathcal{F}_{k,\ell}\left( \eta \right) = \mathcal{F}_{k,\ell}\left( y=0, \eta \right)$. For $m=0$ and $m=1$ Eq. (\ref{eq:approximationyakovlev}) gives the neutrino and anti-neutrino number emissivity (neutrino production rate), and the neutrino and anti-neutrino energy emissivity (energy per unit volume per unit time) for a certain flavour $i$, respectively. Using Eq.~(\ref{eq:approximationyakovlev}) we find that the ratio of emission rates between electronic and non-electronic neutrino flavours obey the relation
\begin{equation}
\frac{\varepsilon^{0}_{e}}{\varepsilon^{0}_{x}} \approx \frac{7}{3}.
\label{eq:neutrinoratio}
\end{equation}
and because of the symmetry of the annihilation process, the neutrinos and anti-neutrinos are produced in equal quantities. We can also find an expression for the average neutrino energy

\begin{equation}
\langle E_{\nu}  \rangle = \langle E_{\bar{\nu}}  \rangle \approx 4.1\,T
\end{equation}\label{eq:neutrinotwomoments}
for all neutrino flavours. The neutrino energy emissivity in Eq.~(\ref{eq:approximationyakovlev}) can be written as
\begin{equation}\label{eq:L_neutrinos}
\epsilon_{e^{-}\!e^{+}} \approx 8.69\times 10^{30}\left(\frac{T}{1\,{\rm MeV}}\right)^9\,\, {\rm MeV}\,{\rm cm}^{-3}\,{\rm s}^{-1},
\end{equation}
which allows us to define an effective neutrino emission region \cite{2018ApJ...852..120B}
\begin{equation}
\Delta r_{\nu} = \frac{\epsilon_{e^{-}\!e^{+}}}{\nabla \epsilon_{e^{-}\!e^{+}}} = \approx 0.08R_{\rm NS}.
\label{neutrinoshell}
\end{equation}
with $R_{\rm NS}$ the radius of the NS. Recollecting results we can make another simplifying assumption~\cite{2018ApJ...852..120B}: Since the neutrino emission region is thin, we will consider it as a spherical shell. This allows us to use the single-angle approximation~\cite{Duan:2006an,Dasgupta:2007ws} and simplify the last term in Eq.~(\ref{eq:Hnu1}). Precisely the \emph{multi-angle} term and the one responsible for kinematic decoherence~\cite{Hannestad:2006nj,Raffelt:2007yz,Fogli:2007bk}. With the single-angle approximation and the inverse square law of flux dilution it is possible to find the explicit dependence in $r$ of each of the potentials in Eq.~(\ref{eq:Hnu1}), namely
\begin{table*}
\begin{adjustbox}{width=2\columnwidth,center}
\begin{tabular}{c c c c c c c c c c c c c c c}
\hline
  $\dot{M}$\T\B & $\rho$ & $T$ & $\eta_{e^{\mp}}$ & $n_{e^{-}}\!-n_{e^{+}}$ & $T_{\nu\bar{\nu}}$ & $\langle E_\nu \rangle$ & $F^{C}_{\nu_e,\bar{\nu}_e}$ & $F^{C}_{\nu_x,\bar{\nu}_x}$ & $n^{C}_{\nu_{e}\bar{\nu}_{e}}$ & $n^{C}_{\nu_{x}\bar{\nu}_{x}}$\ & $\sum_{i}\,n^{C}_{\nu_{i}\bar{\nu}_{i}}$ \\ 
	$(M_\odot$~s$^{-1}$)\B & (g~cm$^{-3})$ & (MeV) &  & (cm$^{-3}$) & (MeV) & (MeV) & (cm$^{-2}$s$^{-1}$) & (cm$^{-2}$s$^{-1}$) & (cm$^{-3})$ & (cm$^{-3})$ & (cm$^{-3})$ \\ \hline
 
     $10^{-6}$\T & $1.12\times10^{7}$ & 2.59 & $\mp 0.193$ & $3.38\times10^{30}$ & 2.93 & 10.61 & $2.40\times 10^{38}$ & $1.03\times 10^{38}$ & $1.60\times10^{28}$ & $6.90\times10^{27}$ & $2.29\times10^{28}$ \\
    $10^{-5}$ & $3.10\times10^{7}$ & 3.34 & $\mp 0.147$ & $9.56\times10^{30}$ & 3.78 & 13.69 & $1.84\times 10^{39}$ & $7.87\times 10^{38}$ & $1.23\times10^{29}$ & $5.20\times10^{28}$ & $1.75\times10^{29}$ \\
    $10^{-4}$ & $8.66\times10^{7}$ & 4.30 & $\mp 0.111$ & $2.61\times10^{31}$ & 4.87 & 17.62 &  $1.39\times 10^{40}$ & $5.94\times 10^{39}$ & $9.24\times10^{29}$ & $3.96\times10^{29}$ & $1.32\times10^{30}$ \\
    $10^{-3}$ & $2.48\times10^{8}$ & 5.54 & $\mp 0.082$ & $7.65\times10^{31}$ & 6.28 & 22.70 & $1.04\times 10^{41}$ & $4.51\times 10^{40}$ & $7.00\times10^{30}$ & $3.00\times10^{30}$ & $1.00\times10^{31}$ \\
  $10^{-2}$ & $7.54\times10^{8}$ & 7.13 & $\mp 0.057$ & $2.27\times10^{32}$ & 8.08 & 29.22 & $7.92\times 10^{41}$  & $3.39\times 10^{41}$ & $5.28\times10^{31}$ & $2.26\times10^{31}$ & $7.54\times10^{31}$ \\ \hline
 
\end{tabular}
\end{adjustbox}
\caption{Characteristics inside the neutrino emission zone for selected values of the accretion rate $\dot{M}$. The symbols $F^{C}$ and $n^{C}$ refer to the neutrino flux and neutrino density at the emission region. The electron fraction is $Y_{e}=0.5$, the pinching parameter for the neutrino spectrum is $\eta_{\nu\bar{\nu}}=2.04$ and the.}
\label{tab:tab1}
\end{table*}
\begin{widetext}
\begin{equation}
\omega_{p,r} =\! \frac{\Delta m^{2}}{2p\langle v_{r} \rangle},\,\, \lambda_{r} \!=\! \sqrt{2}G_{F}\left(n_{e^{-}}\!-n_{e^{+}}\right)\frac{1}{\langle v_{r} \rangle}, \,\, \mu_{r} \!=\!\frac{ \sqrt{2}G_{F}}{2}\left(\sum_{i\in\{e,x\}}\!n^{C}_{\nu_{i}\bar{\nu}_{i}}\right)\left( \frac{R_{\rm NS}}{r} \right)^{2}\left( \frac{1 - \langle v_{r} \rangle^{2}}{\langle v_{r} \rangle} \right),
\label{eq:potentials}
\end{equation}
\end{widetext}
where
\begin{equation}
\langle v_{r} \rangle = \frac{1}{2}\left[ 1+\sqrt{1 - \left(\frac{R_{\rm NS}}{r}\right)^{2}} \right].
\label{eq:averageradialvelocity}
\end{equation}

Using Eq.~(\ref{eq:approximationyakovlev}) and the hydrodynamic simulations in~\cite{2016ApJ...833..107B} we can obtain the thermodynamic properties of the accreting matter at the NS surface (see Table~\ref{tab:tab1}) which in turn are the initial condition to solve Eq.~(\ref{eq:Hnu1}) and obtain an approximate behaviour of oscillations.

\begin{figure*}
\centering
\textbf{\large Inverted Hierarchy}\par\medskip
\includegraphics[width=0.42\hsize,clip]{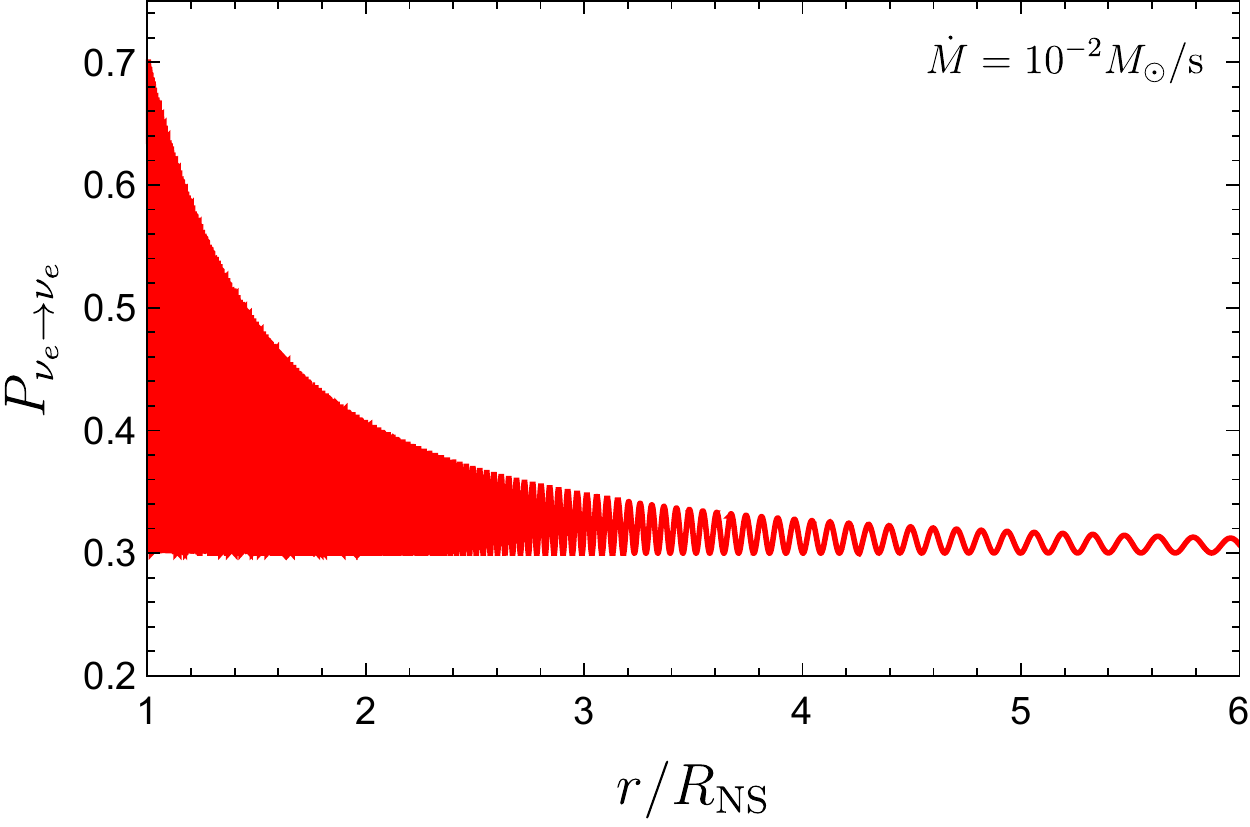}\includegraphics[width=0.42\hsize,clip]{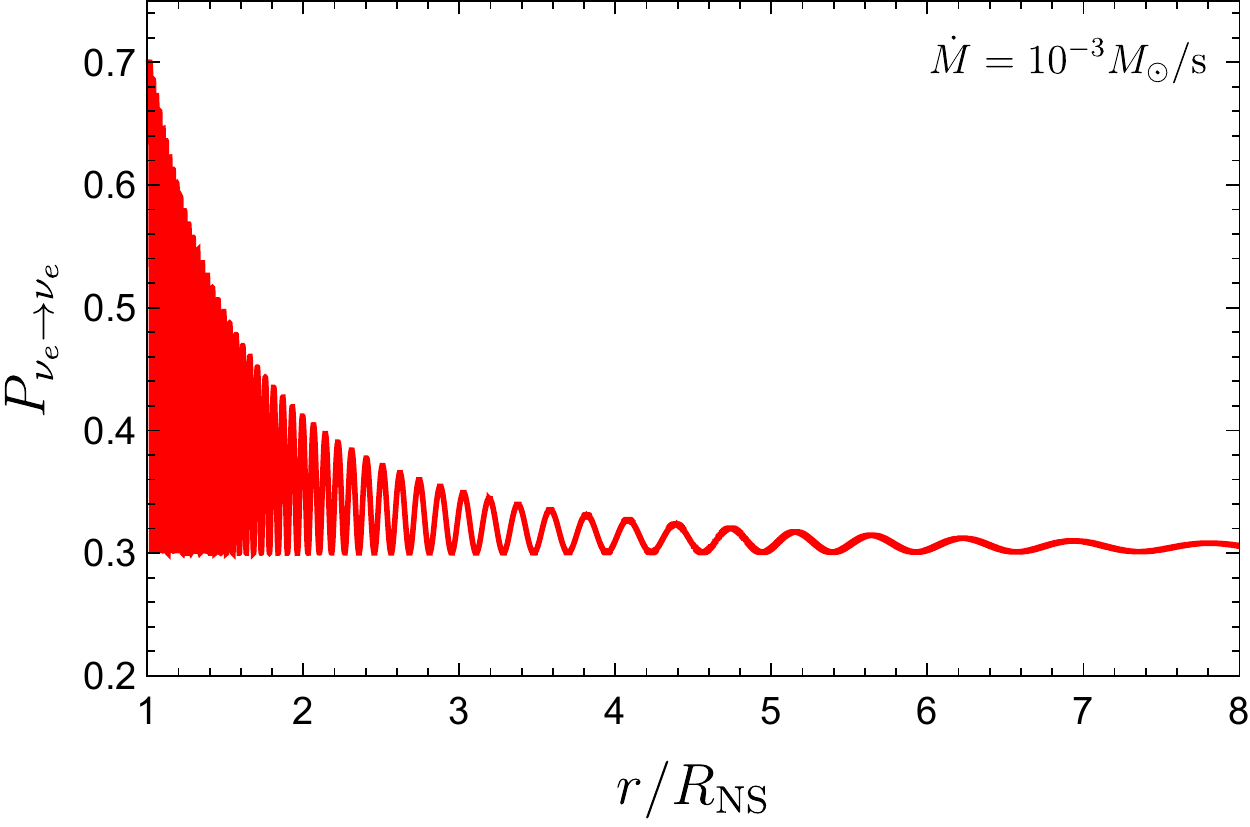}\\
\includegraphics[width=0.42\hsize,clip]{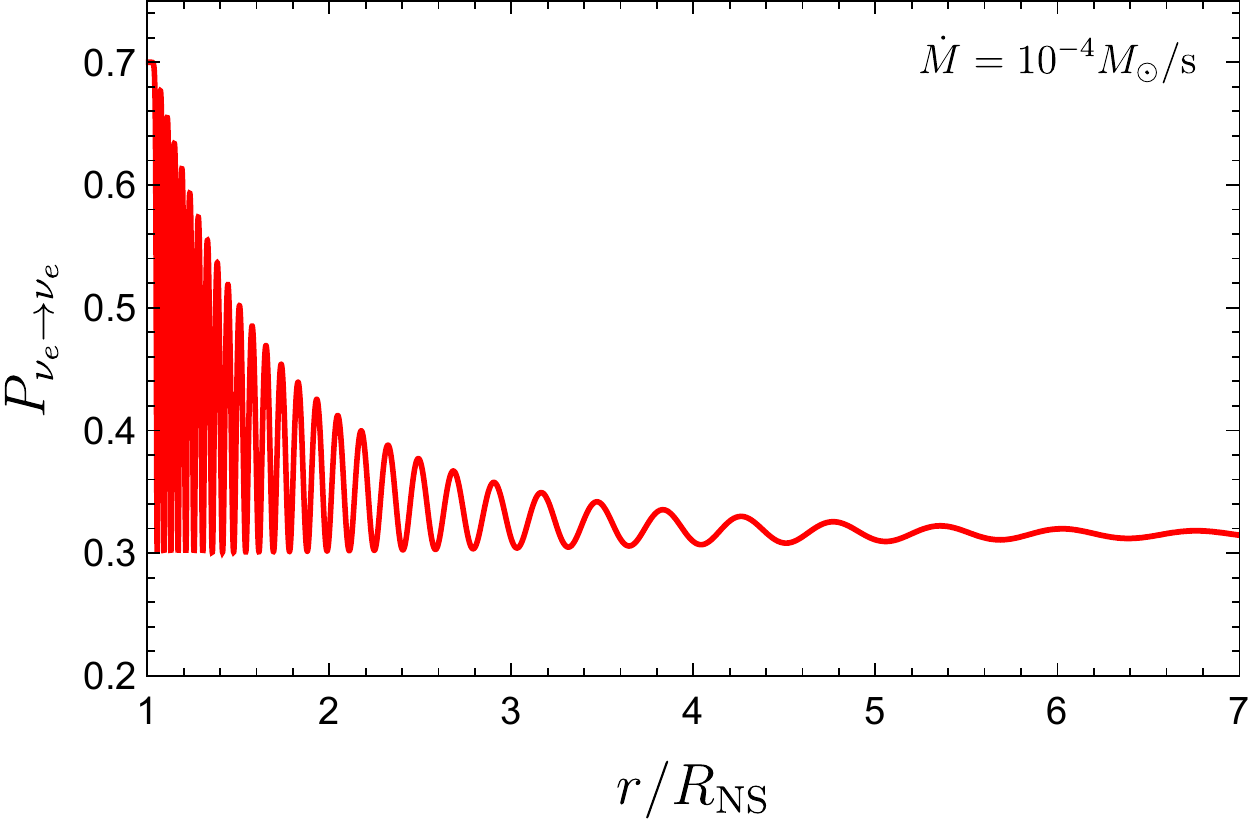}\includegraphics[width=0.42\hsize,clip]{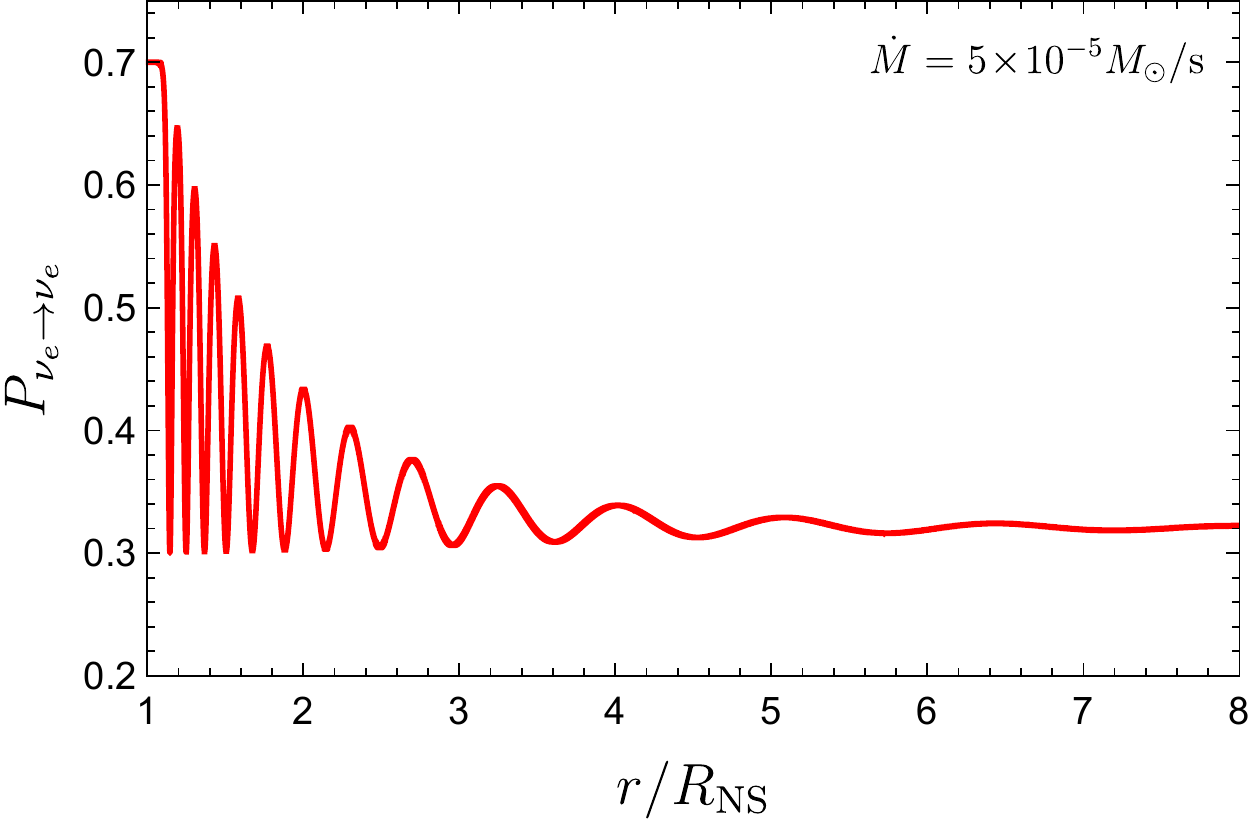}
\caption{Neutrino flavour evolution for inverted hierarchy. Electron neutrino survival probability is shown as a function of the radial distance from the NS surface. The curves for the electron anti-neutrino match the ones for electron neutrinos.} 
\label{fig:singleangle}
\end{figure*}
\begin{figure*}
\centering
\textbf{\large Normal Hierarchy}\par\medskip
\includegraphics[width=0.42\hsize,clip]{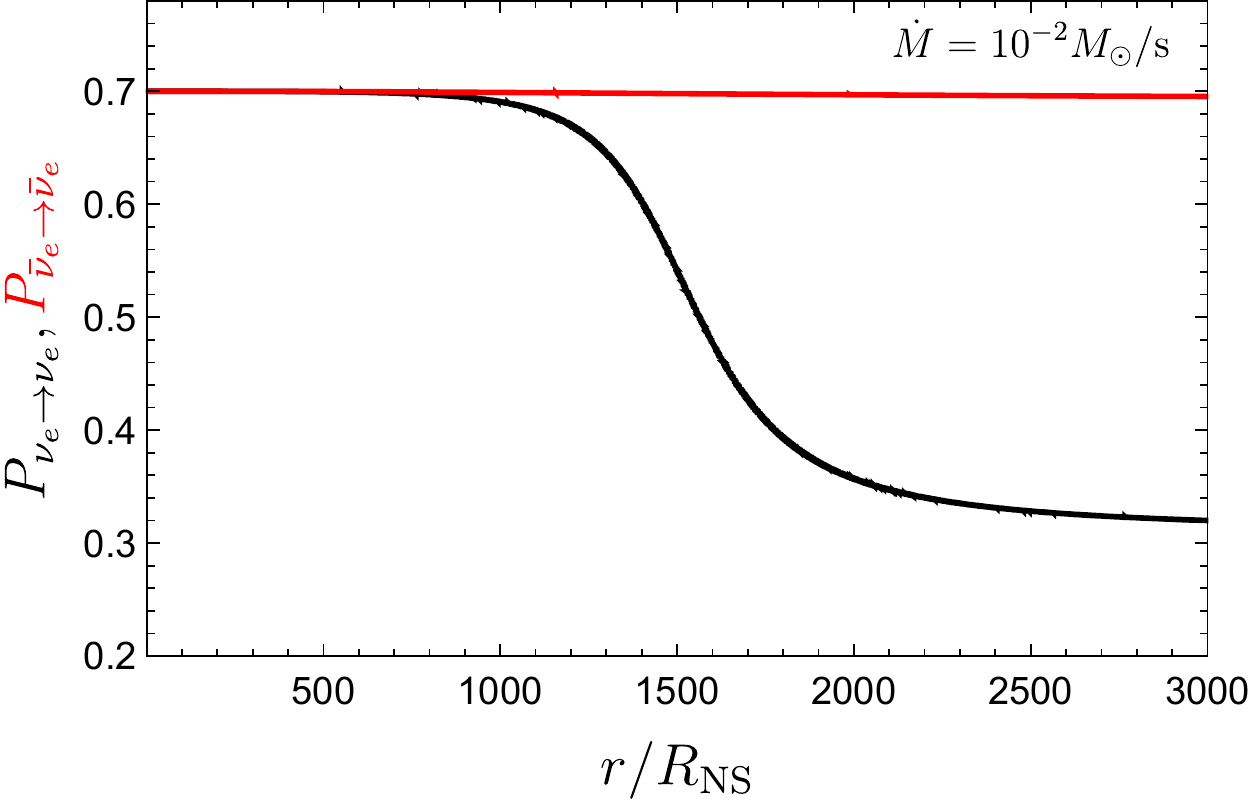}\includegraphics[width=0.42\hsize,clip]{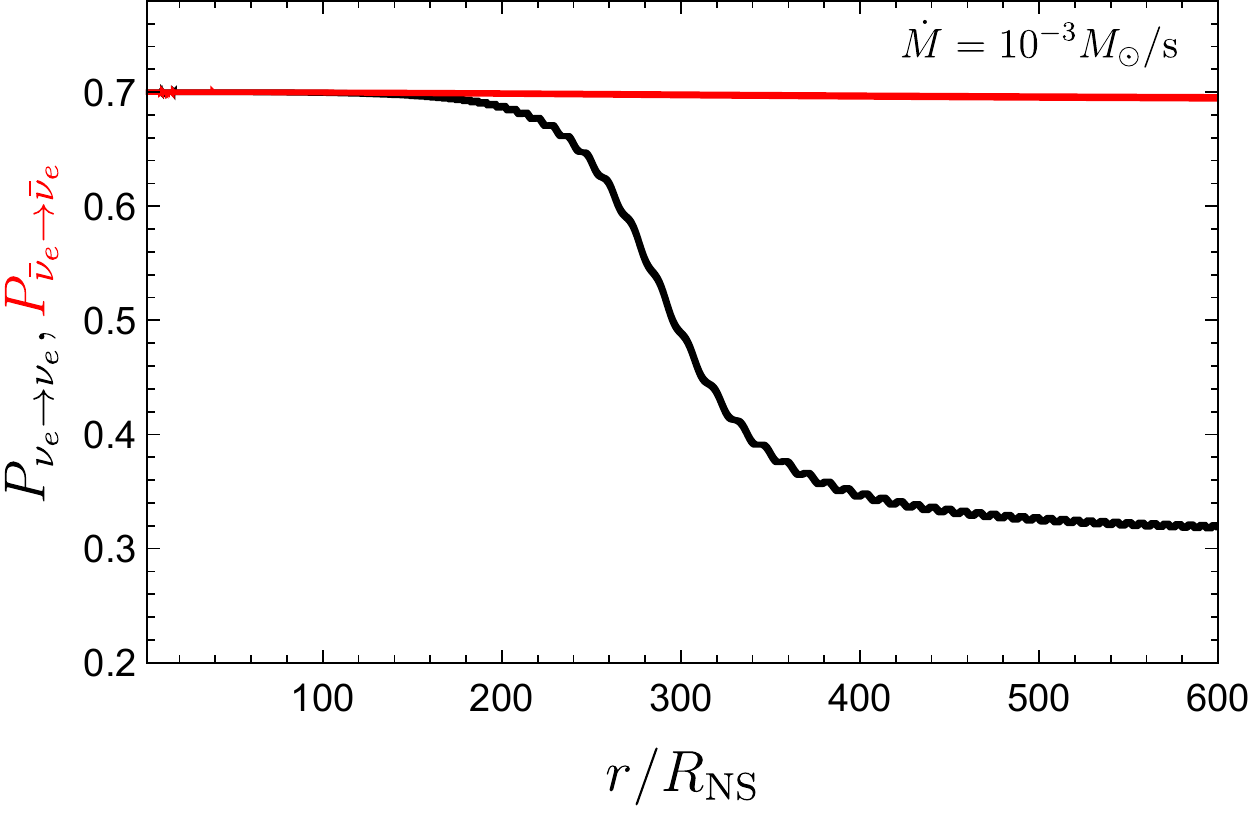}\\
\includegraphics[width=0.42\hsize,clip]{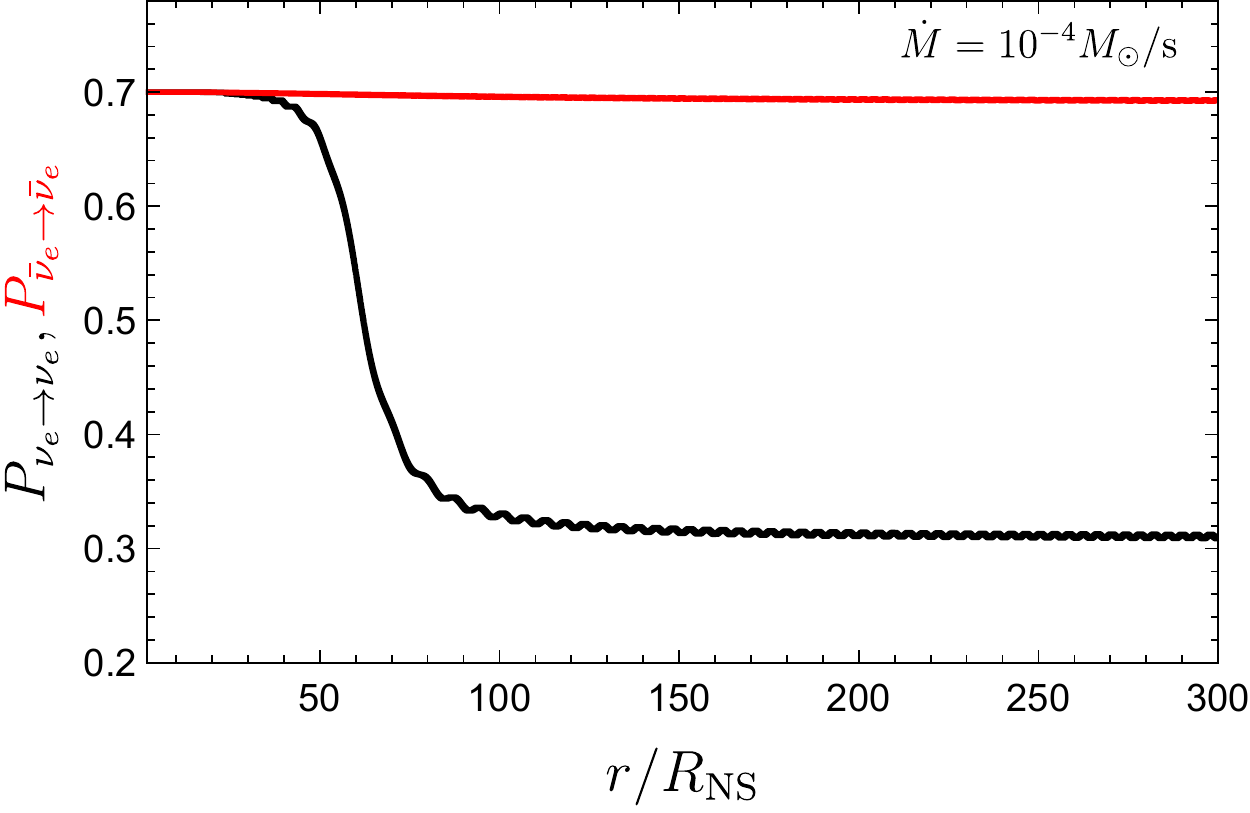}\includegraphics[width=0.42\hsize,clip]{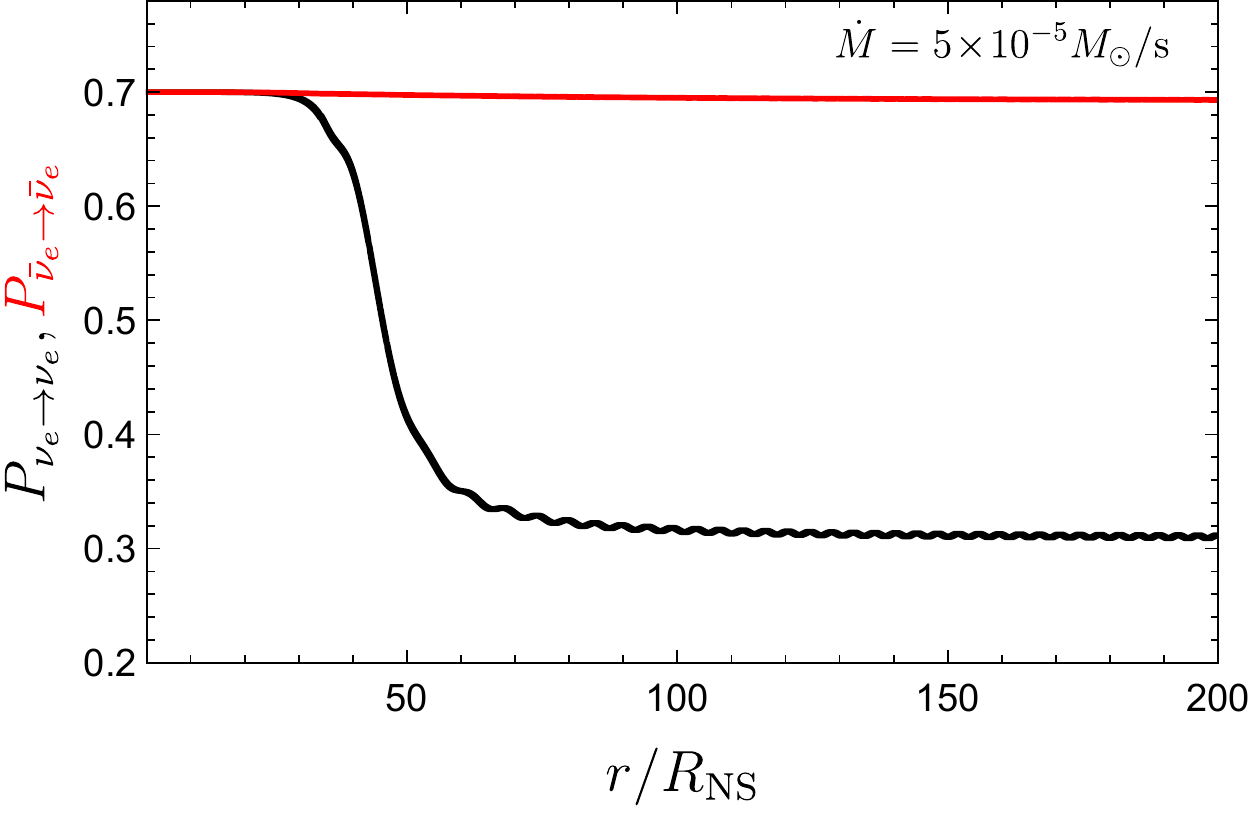}
\caption{Electron neutrino and anti-neutrino flavour evolution for normal hierarchy. The survival probability is shown as a function of the radial distance from the NS surface.} 
\label{fig:singleanglesolutions}
\end{figure*}

In Figs.~\ref{fig:singleangle} and \ref{fig:singleanglesolutions} we show the solution of Eqs.~(\ref{eq:Hnu1}) for both normal and inverted hierarchies using a monochromatic spectrum dominated by the average neutrino energy for $\dot{M}=10^{-2},10^{-3},10^{-4}$ and $5 \times 10^{-5} M_{\odot}$~s$^{-1}$. For the inverted hierarchy, there is no difference between the neutrino and anti-neutrino survival probabilities. This should be expected since for these values of $r$ the matter and self-interaction potentials are much larger than the vacuum potential, and there is virtually no difference between Eqs.~(\ref{eq:Hnu1}). Also, note that the anti-neutrino flavour proportions in Tab.~\ref{tab:tab1} remain virtually unchanged for normal hierarchy while the neutrino flavour proportions change drastically around the point $\lambda_{r} \sim \omega_{r}$. From these solutions we can calculate the oscillation length to be 
\begin{equation}
t_{\rm osc} \approx (0.05-1) \,\,{\rm km}
\label{length}
\end{equation}
which agree with the algebraic estimations in~\cite{Hannestad:2006nj,Raffelt:2007yz}. Clearly, the full equations of oscillations are highly non-linear so the solution may not reflect the real neutrino flavour evolution. Concerning the single-angle approximation, it is discussed in \cite{Hannestad:2006nj,Raffelt:2007yz,Fogli:2007bk} that in the more realistic multi-angle approach, kinematic decoherence happens for both mass hierarchies. And in \cite{EstebanPretel:2007ec} the conditions for decoherence as a function of the neutrino flavour asymmetry have been discussed. It is concluded that if the symmetry of neutrinos and anti-neutrinos is broken beyond the limit of $O(25\%)$, i.e., if the difference between emitted neutrinos and anti-neutrinos is roughly larger than 25\% of the total number of neutrinos in the medium, decoherence becomes a sub-dominant effect. As a direct consequence of the peculiar symmetric situation we are dealing with, in which neutrinos and anti-neutrinos are produced in similar numbers, bipolar oscillations happen and, as we have already discussed, they present very small oscillation length as shown in Eq.~(\ref{length}). Note also that the bipolar oscillation length depends on the neutrino energy. Therefore, the resulting process is equivalent to an averaging over the neutrino energy spectrum and an equipartition among different neutrino flavours is expected~\cite{Raffelt:2007yz}. Although, for simplicity, we are dealing with the two neutrino hypothesis, this behavior is easily extended to the more realistic three neutrino situation. We assume, therefore, that at few kilometers from the emission region neutrino flavour equipartition is a reality:
\begin{equation}
\nu_e:\nu_\mu:\nu_\tau=1:1:1.
\label{eq:proportion}
\end{equation}

After leaving the emission region, beyond $r\approx R_{\rm NS}+\Delta r_{\nu}$, where $\Delta r_{\nu}$ is the width defined in Eq.~(\ref{neutrinoshell}), the effective neutrino density quickly falls in a asymptotic behavior $\mu_{r} \approx 1/r^{4}$. The decay of $\lambda_{r}$ is slower. Hence, very soon the neutrino flavour evolution is determined by the matter potential. Matter suppresses neutrino oscillations and we do not expect significant changes in the neutrino flavour content along a large region. Nevertheless, the matter potential can be so small that there will be a region along the neutrino trajectory in which it can be compared with the neutrino vacuum frequencies and the higher and lower resonant density conditions will be satisfied. Using the results in~\cite{Fogli:2003dw,2018ApJ...852..120B} we can include the matter effects an compare in Tab.~\ref{tab:tabfluxes} the flavour content at the emission region and after decoherence and the MSW resonance.

Finally, we note that for accretion rates $\dot{M} < 5\times\!10^{-5}M_{\odot}$~s$^{-1}$, either the matter potential is close enough to the vacuum potential and the MSW condition is satisfied, or both the self-interaction and matter potentials are so low that the flavor oscillations are only due to the vacuum potential. In both cases, bipolar oscillations are not present~\cite{2018ApJ...852..120B}. Without bipolar oscillations, it is not possible to guarantee that decoherence will be complete and Eq.~(\ref{eq:proportion}) is no longer valid.

\begin{table*}
\begin{adjustbox}{width=2\columnwidth,center}
\begin{tabular}{c c c c c c c c c c c c}
\hline
  & $n^{0}_{\nu_{e}}/n$\T\B & $n^{0}_{\bar{\nu}_{e}}/n$\T\B & $n^{0}_{\nu_{x}}/n$\T\B & $n^{0}_{\bar{\nu}_{x}}/n$\T\B & $n_{\nu_{e}}/n$\T\B & $n_{\bar{\nu}_{e}}/n$\T\B & $n_{\nu_{x}}/n$\T\B & $n_{\bar{\nu}_{x}}/n$\T\B \\ 
 \hline\hline
    Normal Hierarchy\T\B & $\frac{1}{6}$\T & $\frac{1}{6}$\T & $\frac{1}{3}$\T & $\frac{1}{3}$\T & $\frac{1}{3}$\T & $\frac{1}{6} + \frac{1}{6}\sin^{2}\theta_{12}$\T & $\frac{1}{6}$\T & $\frac{1}{3}-\frac{1}{6}\sin^{2}\theta_{12}$\T \\ \hline
    Inverted Hierarchy\T\B & $\frac{1}{6}$\T & $\frac{1}{6}$\T & $\frac{1}{3}$\T & $\frac{1}{3}$\T & $\frac{1}{6} + \frac{1}{6}\cos^{2}\theta_{12}$\T & $\frac{1}{3}$\T & $\frac{1}{3}-\frac{1}{6}\cos^{2}\theta_{12}$\T & $\frac{1}{6}$\T \\ \hline    
\end{tabular}
\end{adjustbox}
\caption{Fraction of neutrinos and anti-neutrinos for each flavour after decoherence and matter effects. $n=2\sum_{i}n_{\nu_{i}}$.}
\label{tab:tabfluxes}
\end{table*}

\subsection{Neutrino Oscillations in Accretion Disks}\label{sec2.2}

In the same BdHN scenario of Sec.~\ref{sec2}, part of the SN ejecta keeps bound to the newborn Kerr BH, forming an accretion disk~\cite{2019ApJ...871...14B}. In order to study analytically the properties of accretion disks, different models make approximations that allow casting the physics of an accretion disk as a two- or even one-dimensional problem. Here, we will consider neutrino-cooled accretion disks (NCADs) which are steady-state~\cite{2019arXiv190901841U,1974ApJ...191..499P}, axisymmetric, thin, alpha-disks models with the following parameters: $\dot{M}$ the accretion rate, $\alpha$ the alpha-viscosity and $a$ the spin of the BH~\cite{2019arXiv190901841U,1973A&A....24..337S,1973blho.conf..343N,1974ApJ...191..499P,1999agnc.book.....K,1999tbha.book.....A,2007ApJ...657..383C,LIU20171}. The procedure to analyse the dynamics of oscillations is similar to the one in~\ref{sec2.1}. The first step is find the neutrino flavour distributions to establish the initial conditions for Eq.~(\ref{eq:Hnu1}), then we have to find each of the potentials and finally solve the equation. To do this we first solve the hydrodynamic model in the absence of oscillations.

\begin{figure*}
\centering
\includegraphics[width=0.315\hsize,clip]{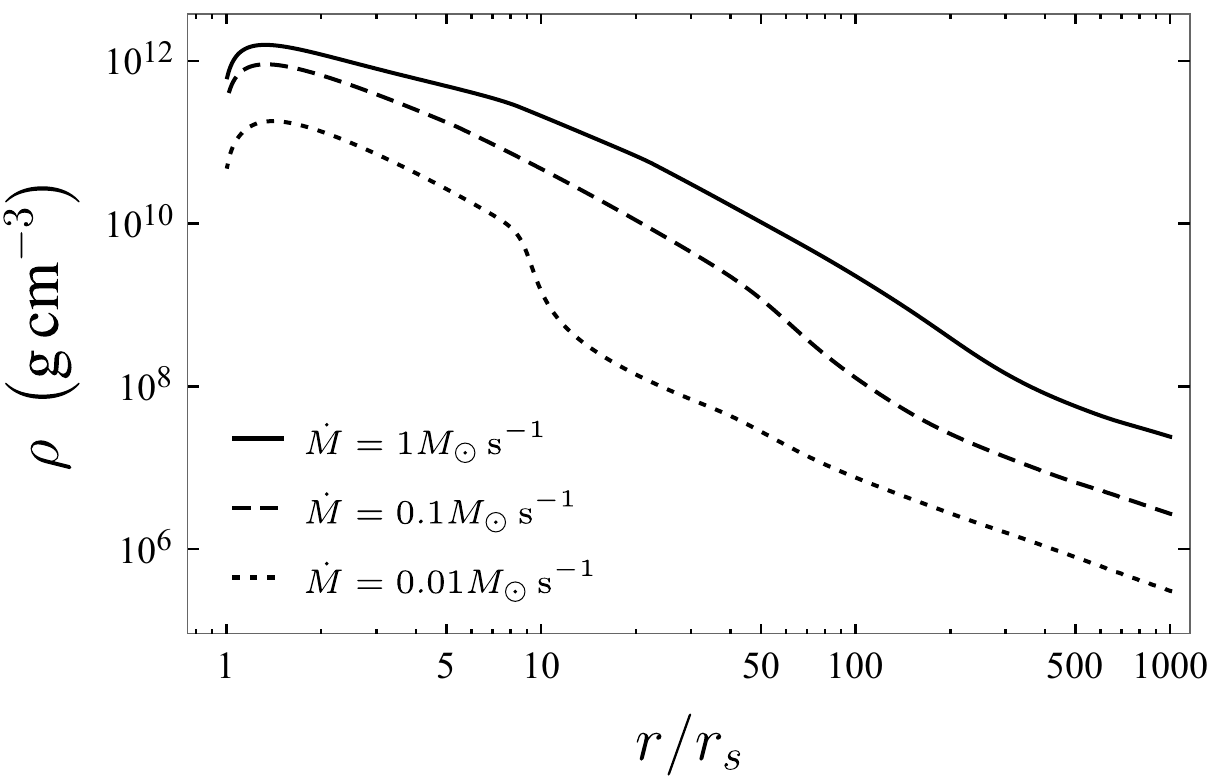}\includegraphics[width=0.315\hsize,clip]{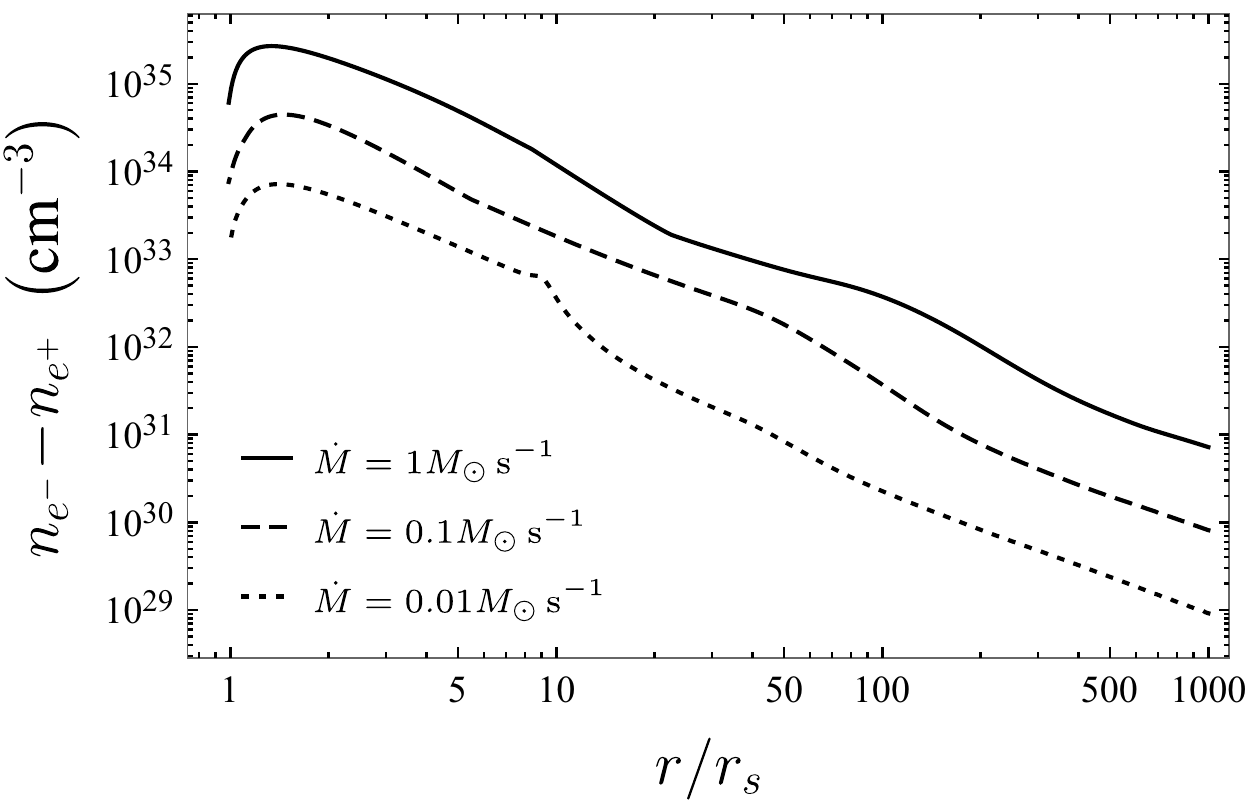}\includegraphics[width=0.315\hsize,clip]{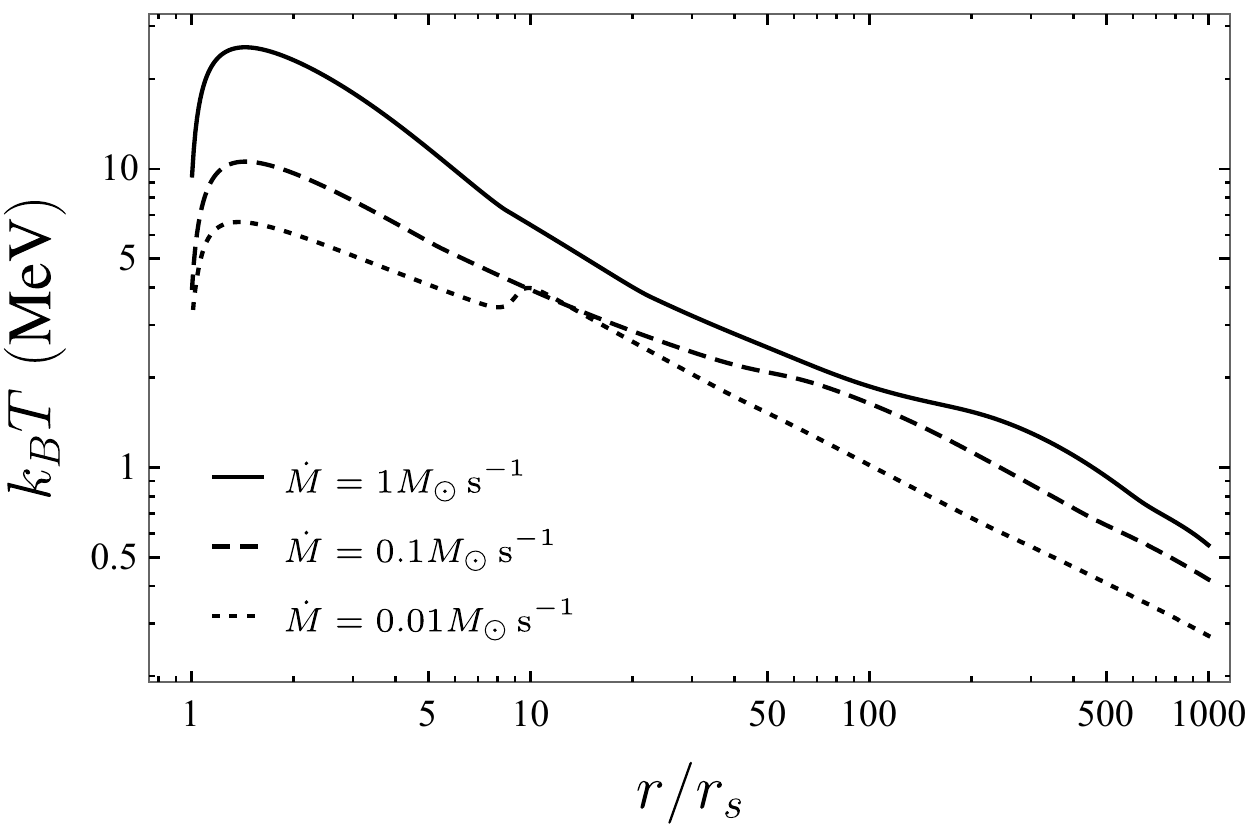}\\
\includegraphics[width=0.315\hsize,clip]{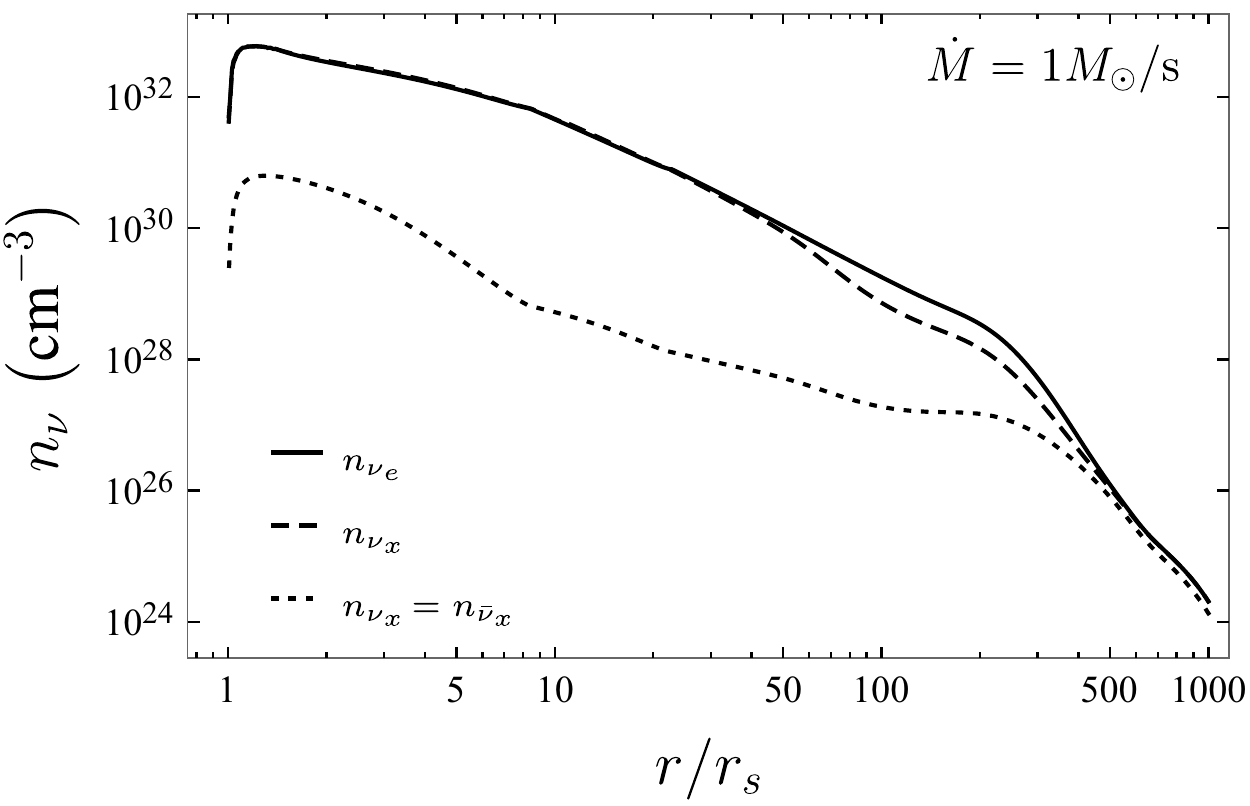}\includegraphics[width=0.315\hsize,clip]{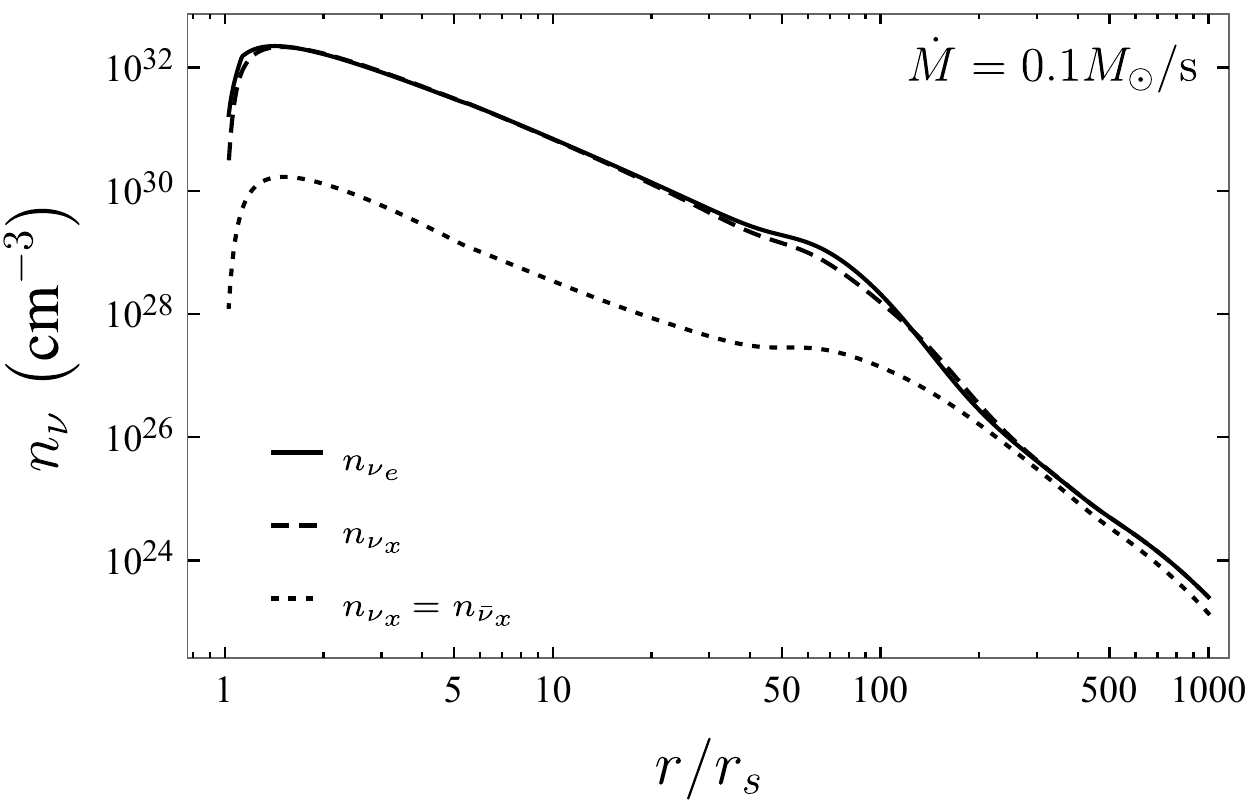}\includegraphics[width=0.315\hsize,clip]{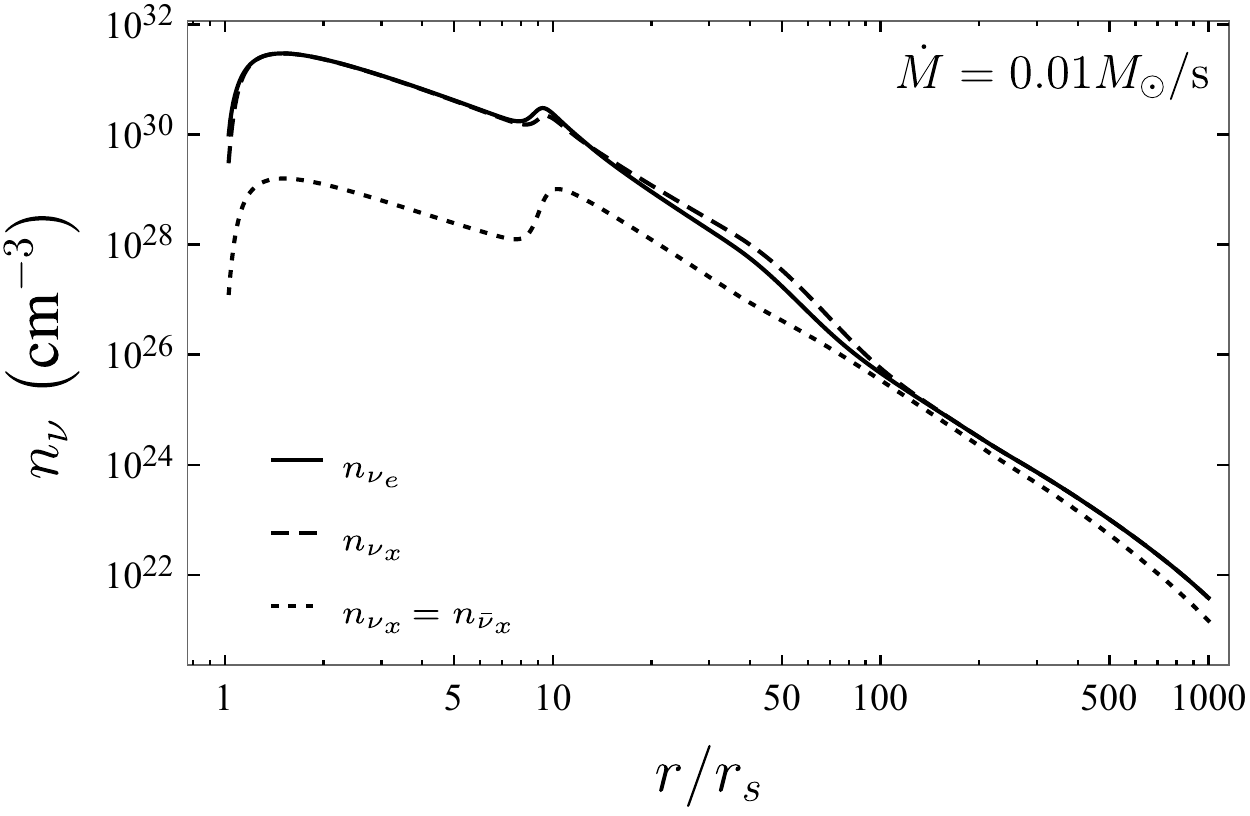}
\includegraphics[width=0.315\hsize,clip]{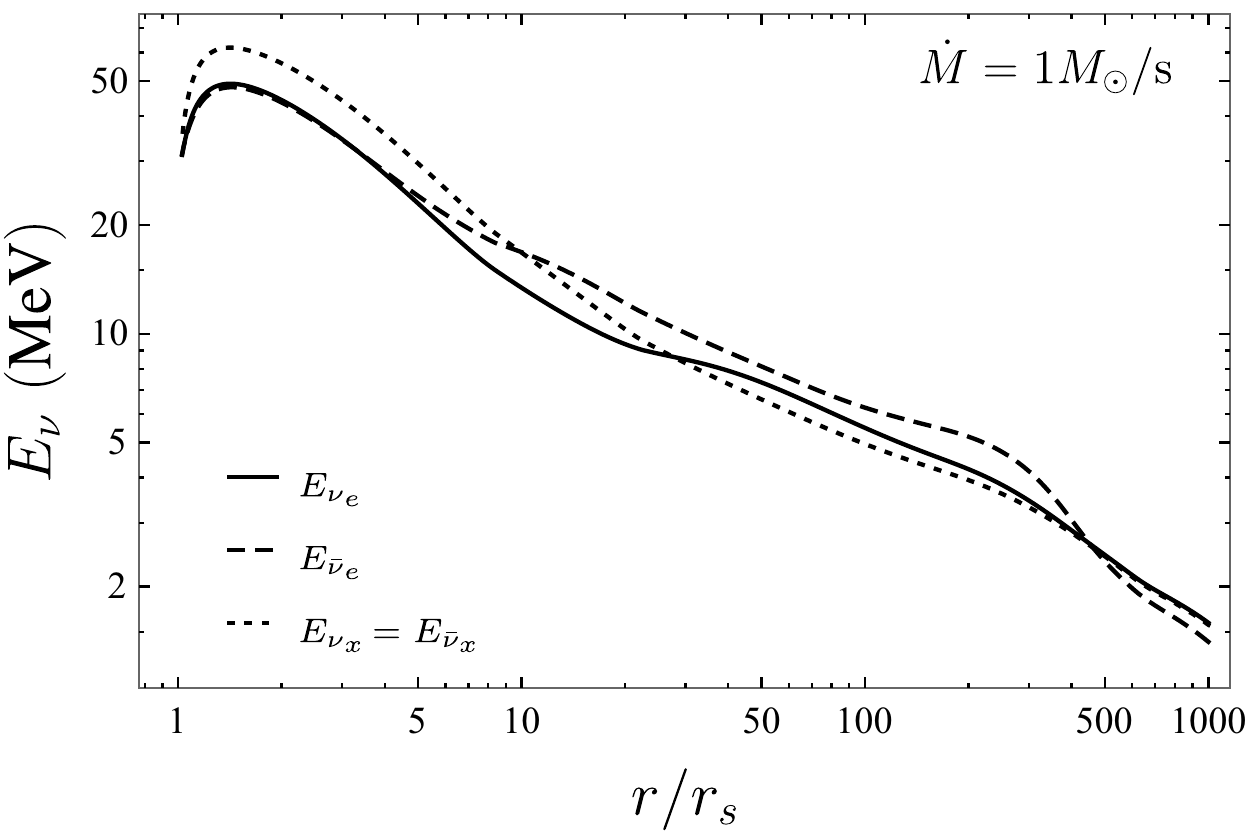}\includegraphics[width=0.315\hsize,clip]{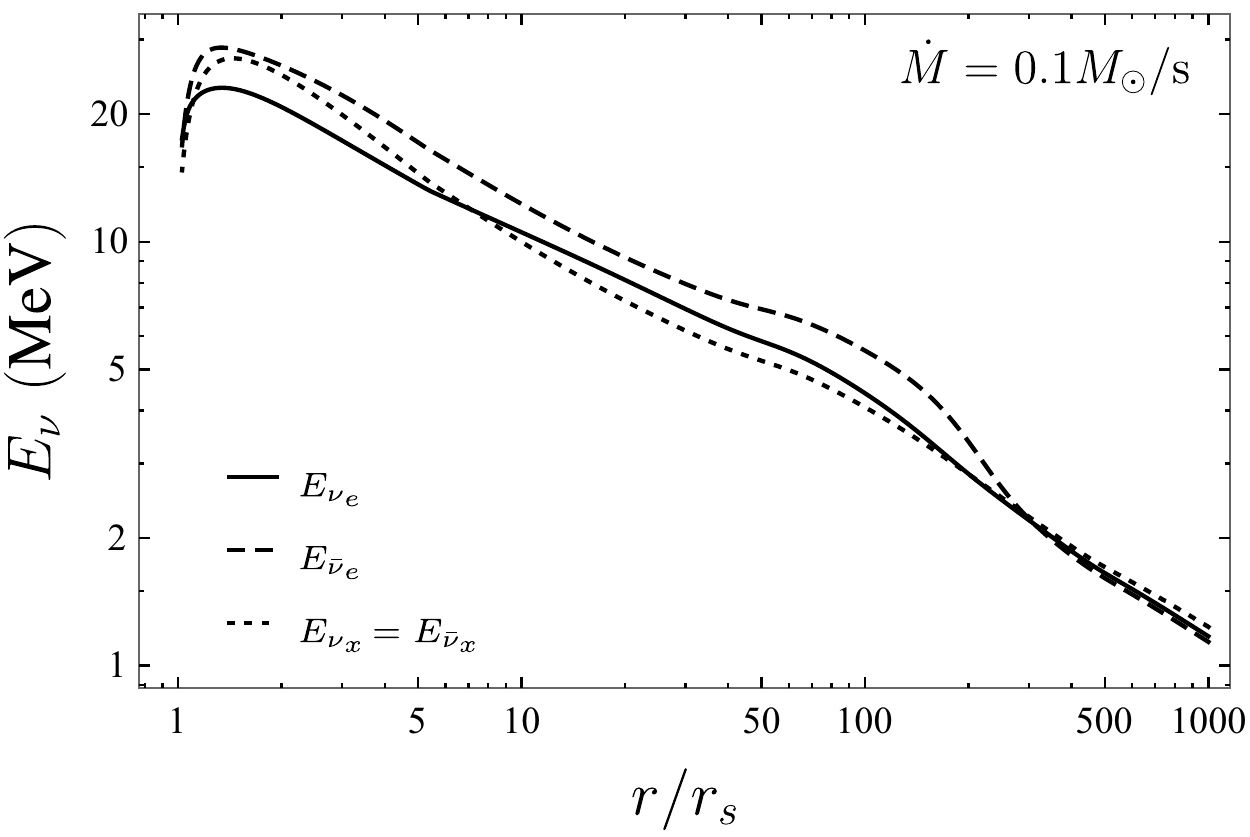}\includegraphics[width=0.315\hsize,clip]{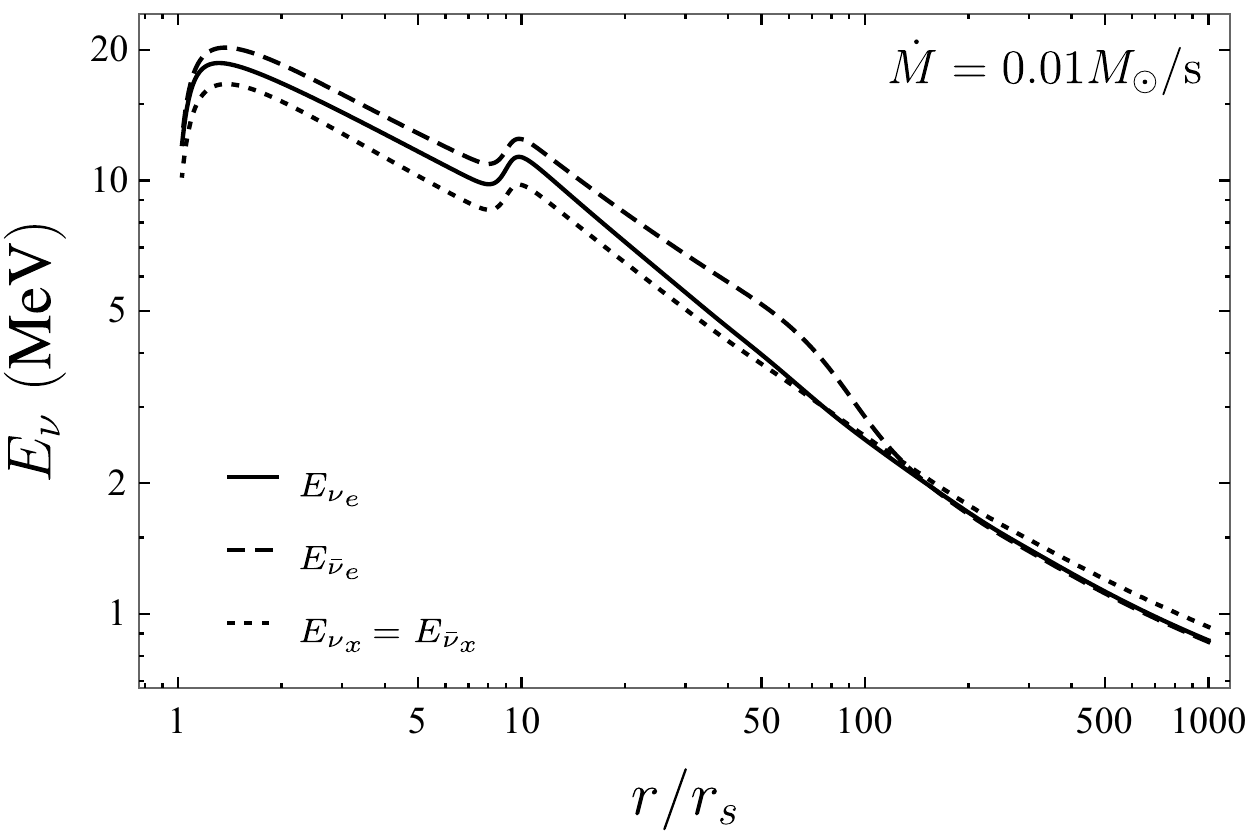}\caption{Properties of accretion disks in the absence of oscillations with $M=3M_{\odot}$, $\alpha = 0.01$, $a = 0.95$ for accretion rates $\dot{M} = 1M_{\odot}$ s$ ^{-1}$, $\dot{M} = 0.1M_{\odot}$ s$ ^{-1}$ and $\dot{M} = 0.01M_{\odot}$ s$ ^{-1}$, respectively.} 
\label{fig:Disks}
\end{figure*}

In Fig.~\ref{fig:Disks} we find the neutrino number densities and energies inside the disk. Note that, as in Sec.~\ref{sec2.1}, the energies of neutrinos are comparable to the ones in spherical accretion and the number of neutrinos and anti-neutrinos are equal. There is also a significant excess of electron neutrinos over non-electron neutrinos. However, there are several key differences that make the analysis in accretion disks more complex. First, an accretion disk has an effective thickness $H$ and neutrinos can be produced at any point inside the disk. This means that it is not possible to set a \emph{surface of emission} as before and due to the lack of spherical symmetry does not allow to use the single-angle approximation. second, close to the BH the effects of curvature may not be negligible, implying that in Eq.~(\ref{eq:Liouville}), when applying the Liouville operator, a term proportional rate of change of the energy of neutrinos $\dot{\mathbf{p}}$ may be present. To simplify the equations of oscillation we consider the local rest frame of the disk (see \cite{1972ApJ...178..347B,1974ApJ...191..507T} for details) and make a set of assumptions:

\begin{figure*}
\centering
\includegraphics[width=0.9\hsize,clip]{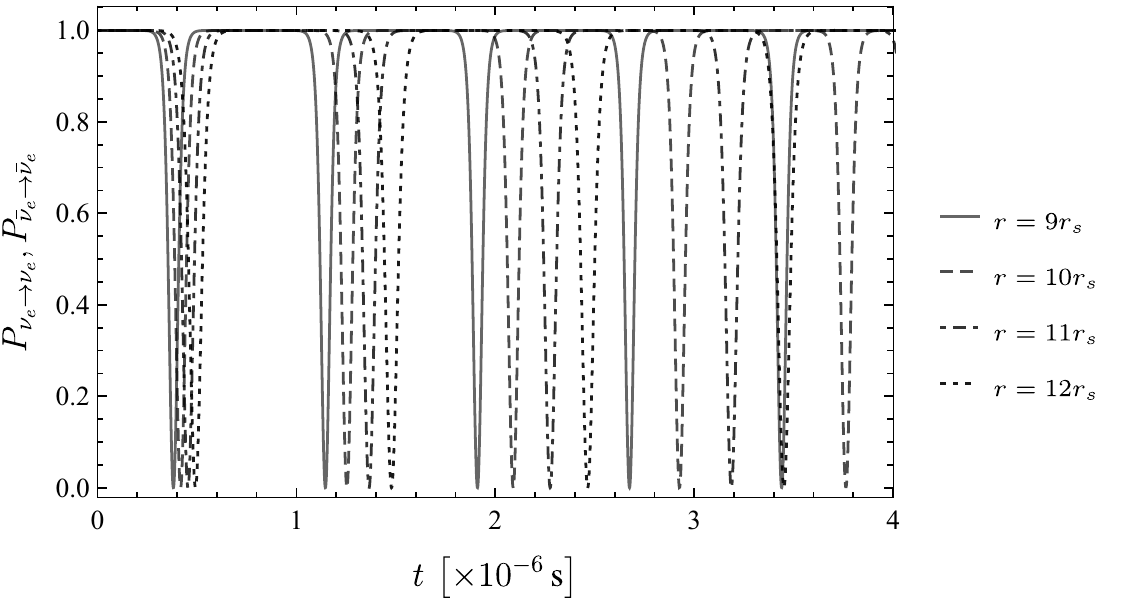}
\caption{Survival provability for electron neutrinos and anti-neutrinos for the accretion disk with $\dot{M}=0.1M_{\odot}$ s$ ^{-1}$ at $r=9r_{s},10r_{s},11r_{s},12r_{s}$.}
\label{fig:SurvSev}
\end{figure*}
\begin{figure*}
\centering
\includegraphics[width=0.49\hsize,clip]{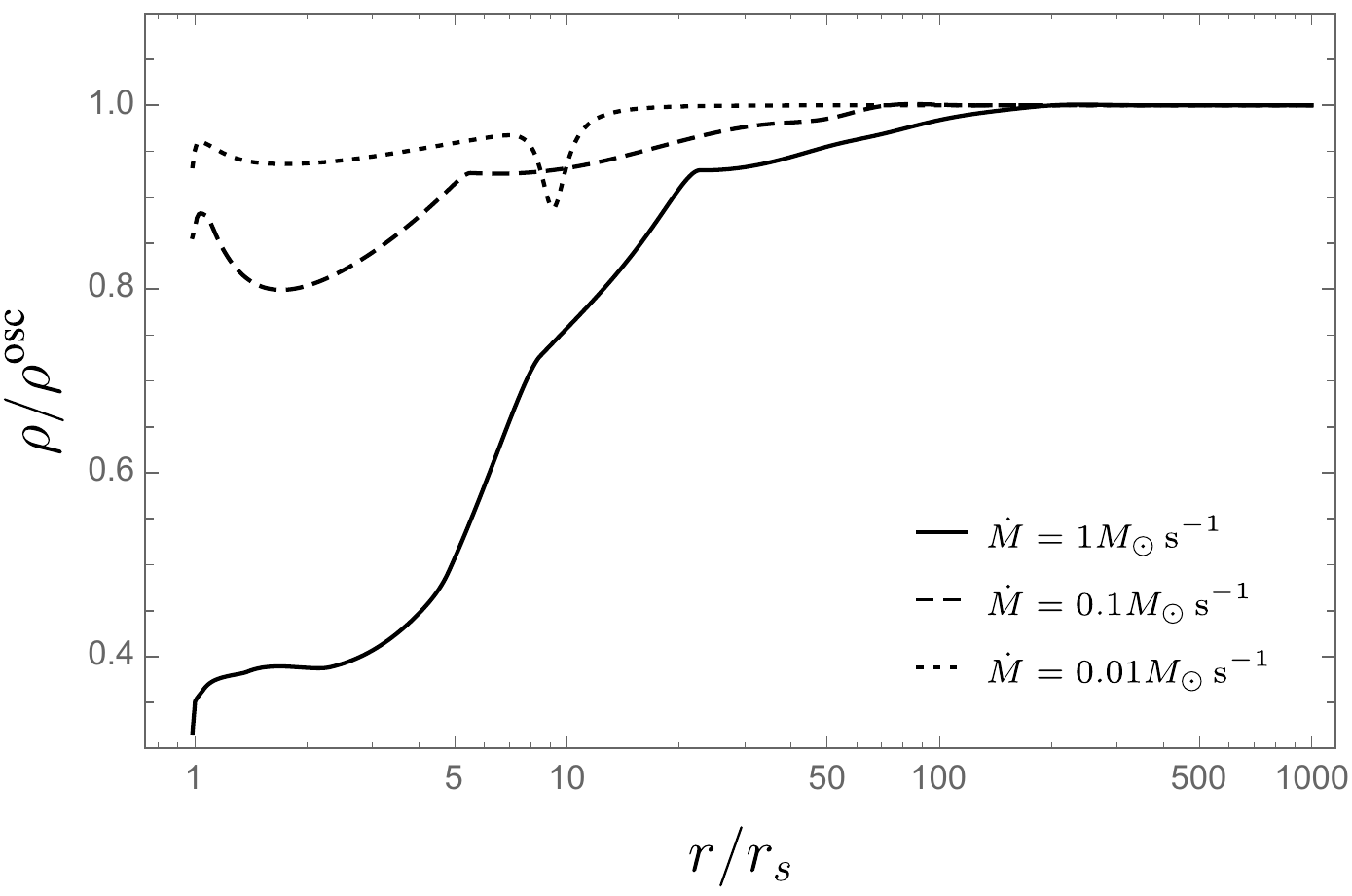}\includegraphics[width=0.49\hsize,clip]{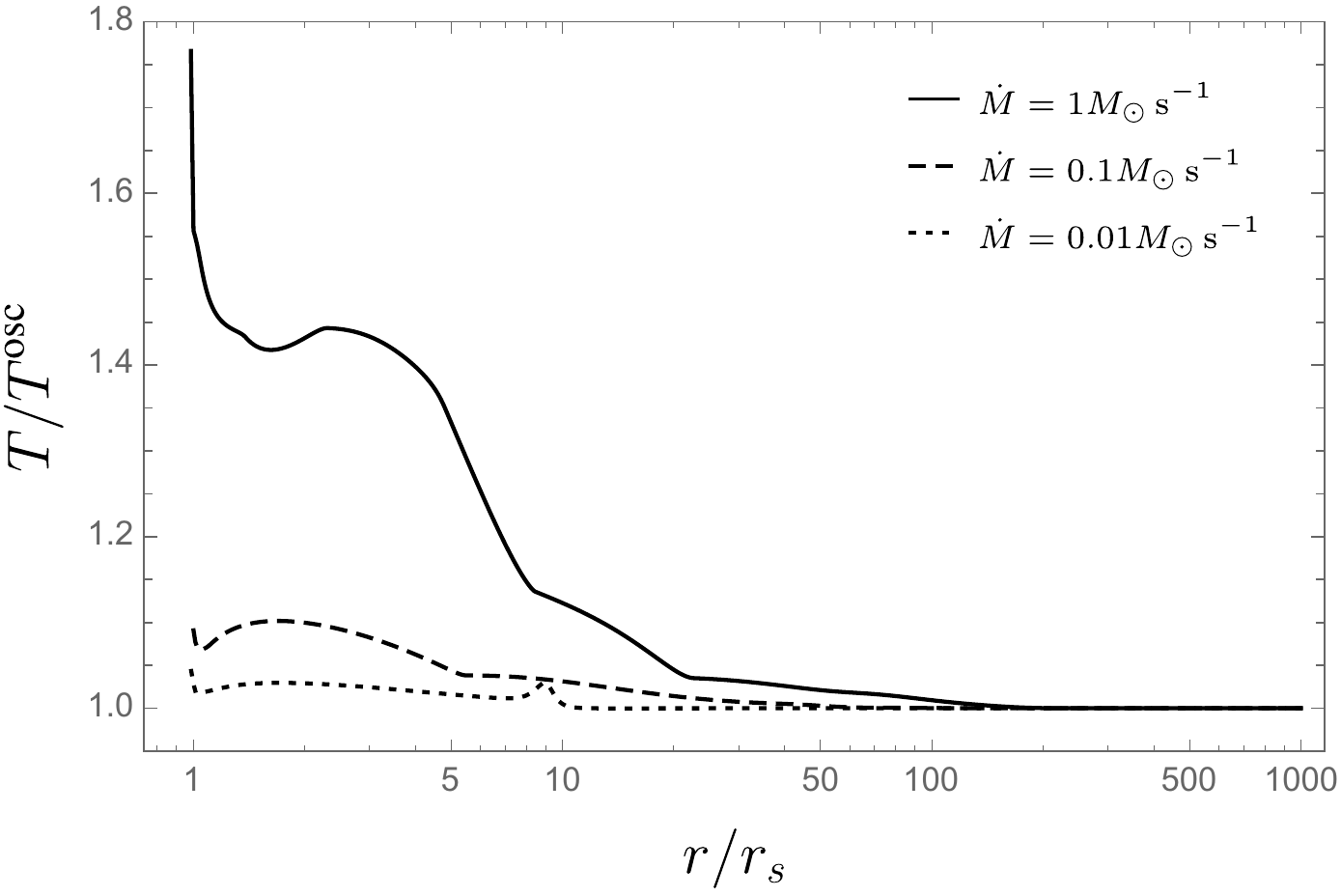}
\caption{Comparison of density and temperature between thin disks with and without neutrino flavour equipartition for selected accretion rates.}
\label{fig:comp}
\end{figure*}

\begin{enumerate}[i]
    
    \item Due to axial symmetry, the neutrino density is constant along the $\mathbf{z}$ direction. Moreover, since neutrinos follow null geodesics, we can set $\dot{p}_{z} \approx \dot{p}_{\phi}=0$. Also, within the thin disk approximation, the neutrino and matter densities are constant along the $\mathbf{y}$ direction and the momentum change due to curvature along this direction can be neglected, that is, $\dot{p}_{y} \approx 0$.
    
    \item In the local rest frame of the disk, the normalized radial momentum of a neutrino can be written as $p_{x} = \pm \frac{r}{\sqrt{r^2-2Mr+M^2a^2}}$ (see \cite{2019arXiv190901841U} for details). Hence, the typical scale of the change of momentum with radius is $\Delta r_{p_{x},\text{eff}} = \left\vert \frac{d\ln{p_{x}}}{dr}  \right\vert^{-1} = \frac{r\left(r^2 -2Mr+M^2a^2\right)}{M\left(Ma^2 - r\right)}$, which obeys $\Delta r_{p_{x},\text{eff}} > r_{s}$ for $r > 2 r_{\text{in}}$. This means that we can assume $\dot{p}_{x} \approx 0$ up to regions very close to the inner edge of the disk. 
    
    \item We define an effective distance $\Delta r_{\rho,\text{eff}} = \left\vert\frac{d \ln\left(n_{e^{-}}-n_{e^{+}} \right)}{dr}\right\vert^{-1}$. For all the systems we evaluated we found that is comparable to the height of the disk $(\Delta r_{\rho,\text{eff}}\sim 2-5$~$r_{s}$). This means that at any point of the disk we can calculate neutrino oscillations in a small regions assuming that both the electron density and neutrino densities are constant.  
    
    \item We neglect energy and momentum transport between different regions of the disk by neutrinos that are recaptured by the disk due to curvature. This assumption is reasonable except for regions very close to the BH but is consistent with the thin disk model (see, e.g.,\cite{1974ApJ...191..499P}). We also assume initially that the neutrino content of neighbouring regions of the disk (different values of $r$) do not affect each other. As a consequence of the results discussed above, we assume that at any point inside the disk and at any instant of time an observer in the local rest frame can describe both the charged leptons and neutrinos as isotropic gases around small enough regions of the disk.

\end{enumerate}

\begin{figure*}
\centering
\includegraphics[width=0.49\hsize,clip]{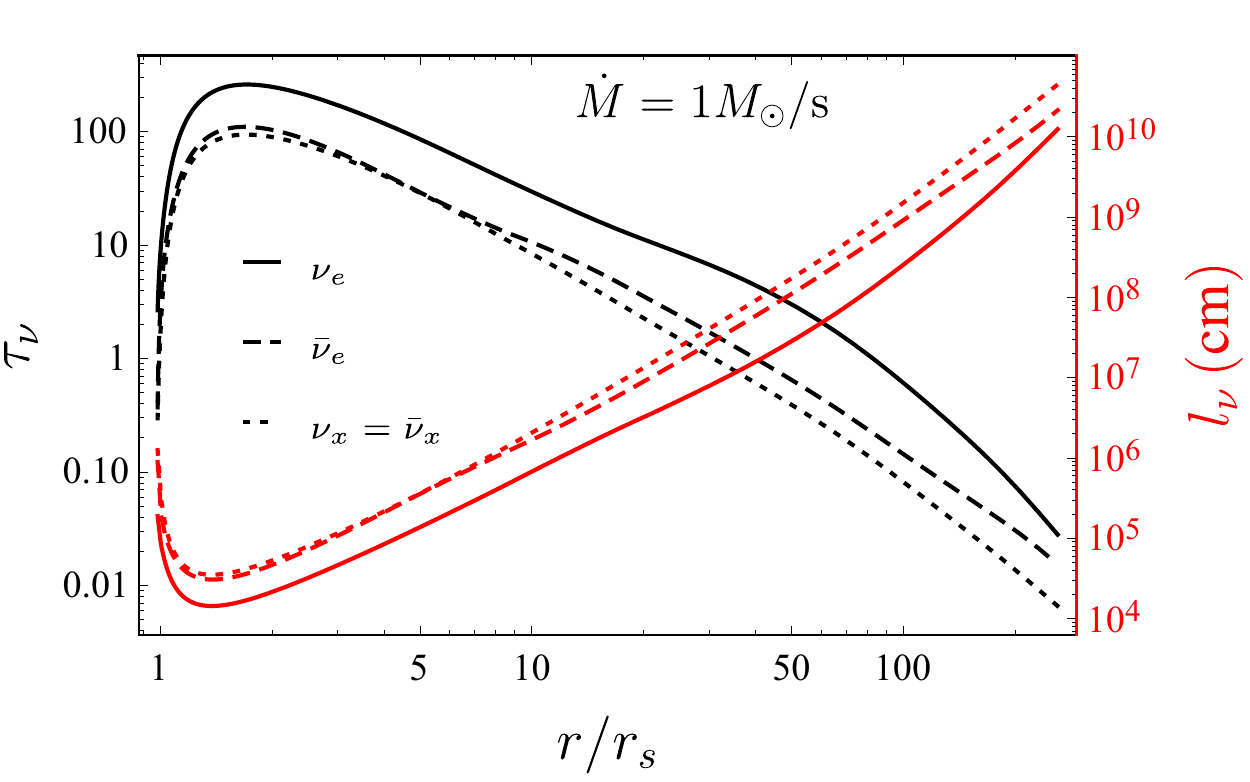}
\includegraphics[width=0.49\hsize,clip]{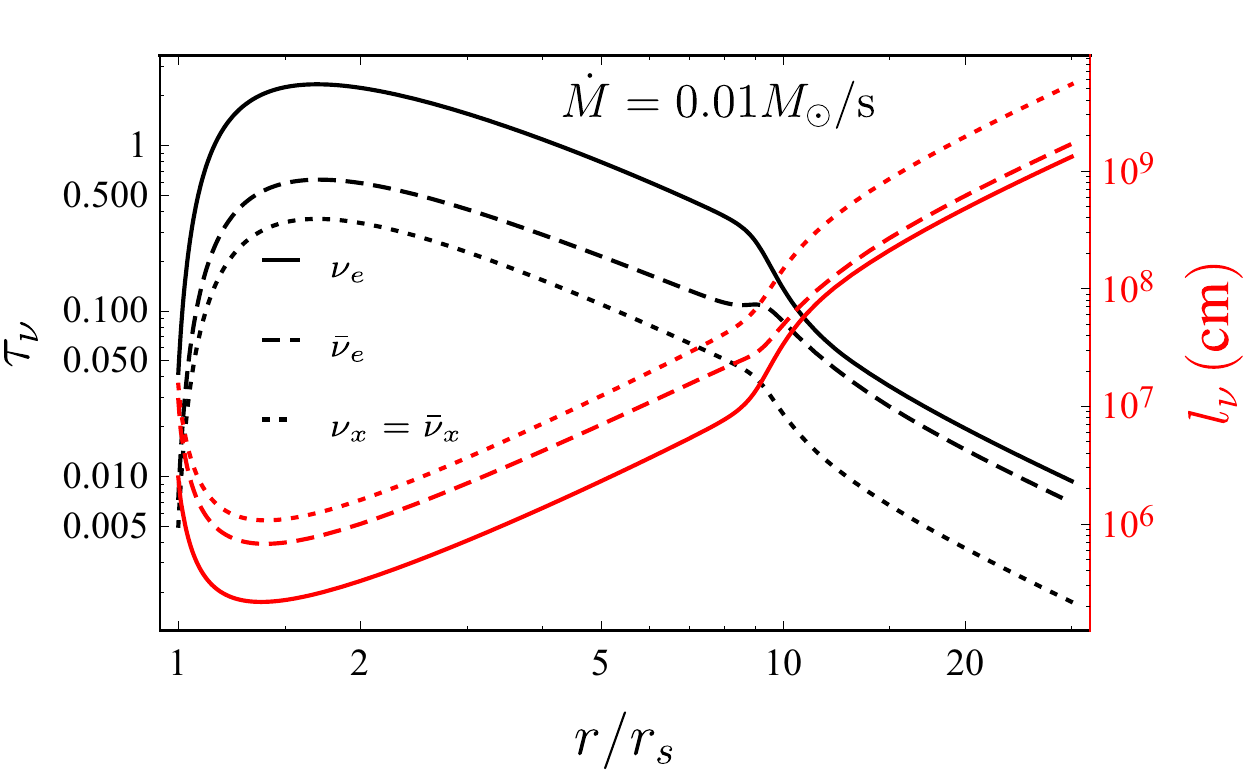}\caption{Total optical depth (left scale) and mean free path (right scale) for neutrinos and anti-neutrinos of both flavours for accretion disks with $\dot{M} = 1M_{\odot}$~s$ ^{-1}$ and $0.01M_{\odot}$~s$ ^{-1}$, between the inner radius and the ignition radius.}
\label{fig:opticaldepth}
\end{figure*}

All assumptions are sensible except iv, which is considerably restrictive. However, we can build our analysis on top of it and use the same results of Sec.~\ref{sec2.1} to generalize the model. Note that with our assumptions, the last term in Eq.~(\ref{eq:Hnu1}) is again simplified. When we calculate the oscillation in different point of the disk (see Fig.~\ref{fig:SurvSev}) we obtain fast flavour transformations with oscillation lengths of the order
\begin{equation}
t_{\rm osc} \approx 10^{-6}\, {\rm s}
    \label{eq:timeosc2}
\end{equation}

Keeping this in mind and given the symmetry between neutrinos and anti-neutrinos in Fig.~\ref{fig:Disks} we note that in~\cite{EstebanPretel:2007ec} it was shown that if the symmetry between neutrinos and anti-neutrinos is not broken beyond the limit of 25\%, kinematic decoherence is still the main effect of neutrino oscillations. Additionally, in \cite{Raffelt:2007yz} it is shown that for asymmetric $\nu\bar{\nu}$ gas, even an infinitesimal anisotropy triggers an exponential evolution towards equipartition. Decoherence happens within a few oscillation cycles of oscillation so we can expect an steady-state, thin disk model to achieve flavour equipartition and is the result of a non-vanishing flux term (which is present in accretion disks due to the increasing density towards the BH) such that at any point, (anti)-neutrinos travelling in different directions, do not experience the same self-interaction potential due to the multi-angle term in the integral of Eq.~(\ref{eq:FullHam}). This effect is of the neutrino mass hierarchy and neutrino flavour equipartition is achieved for both hierarchies. Within the disk dynamic, this is equivalent to imposing the condition
\begin{equation}
    \langle P_{\nu_e \to \nu_e} \rangle= \langle P_{\bar{\nu}_e \to \bar{\nu}_e} \rangle = 0.5.
\end{equation}\label{eq:equipartition}

Within this condition, we can compare the behaviour of disks with and without flavour equipartition. Figure~\ref{fig:comp} shows that equipartition increases the disk density and reduces the temperature where the neutrino emission is important. The effect is mild for low accretion rates while very pronounced for high ones. Thus result can be explained as follows: for low accretion rates the neutrino optical depth for all flavors is $\tau_{\nu\bar{\nu}} \lesssim 1$ (see Fig.~\ref{fig:opticaldepth}), hence neutrinos, regardless of their flavour, are free to leave the disk. When the initial (mainly electron flavour) is redistributed among both flavours, the total neutrino cooling remains virtually unchanged and the disk evolves as if equipartition had never occurred, save the new emission flavour content. On the other hand, when accretion rates are high, the optical depth obeys $\tau_{\nu_{x}}\approx\tau_{\bar{\nu}_{x}}\lesssim\tau_{\bar{\nu}_{e}} < \tau_{\nu} \sim 10^3$. The $\nu_{e}$ cooling is more heavily suppressed than the others. When flavours are redistributed, the \textit{new} $\nu_{x}$ particles a free to escape, enhancing the total cooling with a consequent reduction of the temperature. As the temperature decreases, a lower internal energy allows for a higher matter density. The net impact of flavour equipartition is to make the disk evolution less sensitive to $\nu_{e}$ opacity. It can be shown (see \cite{2019arXiv190901841U} for details) that it increases the total cooling efficiency by the precise factor
\begin{equation}
\frac{1}{2}\left(1 + \frac{\langle E_{\nu_{x}} \rangle}{\langle E_{\nu_{e}} \rangle}\frac{1 + \tau_{\nu_{e}}}{1 + \tau_{\nu_{x}}}\right).
    \label{eq:fluxcomp}
\end{equation}

The main difference with the previous system is that, for similar accretion rates, the density of the accretion disk can be high dense to impede, or even trap neutrinos within it. However, since electron and non-electron neutrinos have different cross sections, the flavour transformations affect not only the dynamics of the disk, but also the neutrino flavour content emerging from the disk. This, in turn, affects the energy deposition rate of the process $\nu+\bar{\nu}\mapsto e^{-}+e^{+}$. In particular, it leads to a deficit of electron neutrinos and a smaller energy deposition rate with respect to previous estimates not accounting by flavour oscillations inside the disk. The exact value of the reduction factor depends on the $\nu_{e}$ and $\nu_{x}$ optical depths but it can be as high as $\sim 5$. We refer the reader to \cite{2019arXiv190901841U} for further details on this subject.

\section{Concluding Remarks}\label{sec3}

We have outlined the implications of neutrino oscillations in two different accreting systems within the BdHN scenario of GRBs. In both, spherical accretion and disk accretion, the emission of neutrinos is a crucial ingredient since they act as the main cooling process that allows the accretion onto the NS (or onto the BH) to proceed at very high rates of up to $1~M_\odot$~s$^{-1}$. Also, the ambient conditions of density and temperature imply the occurrence of neutrino flavour oscillations, with a relevant role of neutrino self-interactions.

We have seen that in spherical accretion the density of neutrinos on top the NS implies that neutrino self-interactions dominate the flavour evolution, leading to collective effects. The latter induce quick flavour conversions with an oscillation length as small as $(0.05$--$1)$~km. Far from the NS surface, the neutrino density decreases and so the matter potential and MSW resonances dominate the flavour oscillations. Owing to the above, the neutrino flavour content emerging from the system is completely different with respect to the one created at the bottom of it, namely on top the NS accreting surface.

Concerning disk accretion onto a BH, we saw that the number densities of electron neutrinos and anti-neutrinos are very similar. As a consequence of this particular environment, very fast pair conversions, $\nu_{e}\bar{\nu}_{e} \rightleftharpoons \nu_{x}\bar{\nu}_{x}$, induced by bipolar oscillations, are obtained for the inverted mass hierarchy case with high oscillation frequencies. However, due to the interaction between neighbouring regions of the disk, the onset of kinematic decoherence with a timescale comparable to the oscillation length induces flavour equipartition among electronic and non-electronic neutrinos throughout the disk. Therefore, the neutrino content emerging from the disk is very different from the one that is usually assumed (see e.g. \cite{2012PhRvD..86h5015M,2016PhRvD..93l3004L}).

Flavour equipartition, while leaving anti-neutrino cooling practically unchanged, it enhances neutrino cooling by allowing the energy contained (and partially trapped inside the disk due to high opacity) within the $\nu_{e}$ gas to escape in the form of $\nu_{x}$, rendering the disk insensible to the electron neutrino opacity. The variation of the flavour content in the emission flux implies a loss in the electron neutrino luminosity and an increase in non-electron neutrino luminosity and $L_{\bar{\nu}_{e}}$. As a consequence, the total energy deposition rate of the process $\nu+\bar{\nu}\to e^{-}+e^{+}$ is reduced. 

These results are only a first step toward the analysis of neutrino oscillations in a novel relativistic astrophysics context that can have an impact on a wide range of astrophysical phenomena: from $e^{-}e^{+}$ plasma production above BHs in GRB models, to r-process nucleosynthesis in disk winds and possible MeV neutrino detectability.

\bibliographystyle{ieeetr}
\bibliography{main}

\end{document}